\documentclass[sigconf]{acmart}
\AtBeginDocument{%
  }

\setcopyright{acmlicensed}
\copyrightyear{2026}
\acmYear{2026}
\acmDOI{XXXXXXX.XXXXXXX}
\acmISBN{978-1-4503-XXXX-X/2018/06}

\usepackage{amsmath} 
\usepackage{amsfonts} 
\usepackage{mathrsfs}
\usepackage{enumitem}
\usepackage{booktabs}  
\usepackage{tcolorbox}
\usepackage{multirow} 
\usepackage{graphicx}  
\usepackage{amsmath}  
\usepackage{mathrsfs}  
\usepackage{subcaption}
\usepackage{caption}
\usepackage{subcaption}

\usepackage{algorithm}
\usepackage{algorithmic}
\allowdisplaybreaks

\begin{document}

% \title{Collaboration and Confrontation: Infusing Personalized Knowledge from Large Language Models into Recommendation}
\title{Synergistic Integration and Discrepancy Resolution of Contextualized Knowledge for Personalized Recommendation}

\author{Lingyu Mu}
% \orcid{0000-0002-7252-5207}
\authornotemark[1]
\affiliation{
  \institution{Institute of Information Engineering, Chinese Academy of Sciences}
  \city{Beijing} 
  \state{} 
  \country{China}
}
\email{mulingyu@iie.ac.cn}

\author{Hao Deng}
\orcid{0009-0002-6335-7405}
\authornote{Contributed equally to this research.} 
\affiliation{%
  \institution{Alibaba International Digital Commerce Group}
   \city{Beijing} 
   \state{} 
   \country{China}
}
\email{denghao.deng@alibaba-inc.com}

\author{Haibo Xing}
\orcid{0009-0006-5786-7627}
\affiliation{%
  \institution{Alibaba International Digital Commerce Group}
  \city{Hangzhou} 
  \state{} 
  \country{China}
}
\email{xinghaibo.xhb@alibaba-inc.com}

\author{Kaican Lin}
\affiliation{%
  \institution{Alibaba International Digital Commerce Group}
  \city{Beijing} 
  \state{} 
  \country{China}
}
\email{linkaican.lkc@alibaba-inc.com}

\author{Zhitong Zhu}
\affiliation{
  \institution{Institute of Information Engineering, Chinese Academy of Sciences}
  \city{Beijing} 
  \state{} 
  \country{China}
}
\email{zhuzhitong@iie.ac.cn}

\author{Yu Zhang}
\orcid{0000-0002-6057-7886}
\affiliation{
  \institution{Alibaba International Digital Commerce Group}
  \city{Beijing} 
  \state{} 
  \country{China}
}
\email{daoji@lazada.com}

\author{Xiaoyi Zeng}
\orcid{0000-0002-3742-4910}
\affiliation{
  \institution{Alibaba International Digital Commerce Group}
  \city{Hangzhou} 
  \state{} 
  \country{China}
}
\email{yuanhan@taobao.com}

\author{Zhengxiao Liu$^\dag$}
\affiliation{
  \institution{Institute of Information Engineering, Chinese Academy of Sciences}
  \city{Beijing} 
  \state{} 
  \country{China}
}
\email{liuzhengxiao@iie.ac.cn}

\author{Zheng Lin}
\authornote{Corresponding authors.}
\affiliation{
  \institution{Institute of Information Engineering, Chinese Academy of Sciences}
  \city{Beijing} 
  \state{} 
  \country{China}
}
\email{linzheng@iie.ac.cn}

\author{Jinxin Hu}
\orcid{0000-0002-7252-5207}
\affiliation{
  \institution{Alibaba International Digital Commerce Group}
  \city{Beijing} 
  \state{} 
  \country{China}
}
\email{jinxin.hjx@alibaba-inc.com}

\renewcommand{\shortauthors}{Trovato et al.}

\begin{abstract}
  The integration of large language models (LLMs) into recommendation systems has revealed promising potential through their capacity to extract world knowledge for enhanced reasoning capabilities. However, current methodologies that adopt static schema-based prompting mechanisms encounter significant limitations: (1) they employ universal template structures that neglect the multi-faceted nature of user preference diversity; (2) they implement superficial alignment between semantic knowledge representations and behavioral feature spaces without achieving comprehensive latent space integration. To address these challenges, we introduce CoCo, an end-to-end framework that dynamically constructs user-specific contextual knowledge embeddings through a dual-mechanism approach. Our method realizes profound integration of semantic and behavioral latent dimensions via adaptive knowledge fusion and contradiction resolution modules. Experimental evaluations across diverse benchmark datasets and an enterprise-level e-commerce platform demonstrate CoCo's superiority, achieving a maximum 8.58\% improvement over seven cutting-edge methods in recommendation accuracy. The framework's deployment on a production advertising system resulted in a 1.91\% sales growth, validating its practical effectiveness. With its modular design and model-agnostic architecture, CoCo provides a versatile solution for next-generation recommendation systems requiring both knowledge-enhanced reasoning and personalized adaptation.
\end{abstract}


\vspace{-3pt}
\vspace{-3pt}


\maketitle
% \noindent\textbf{Relevance to WWW'26:} This work advances user modeling for Web-scale personalization using a collaboration-contradiction mechanism to deeply align Large Language Model knowledge with user behavior on the Web, proven effective on large-scale online platforms.
% \maketitle
% \begin{quote} % The 'quote' environment creates a visually distinct, indented block.
% \textbf{Relevance to WWW'26.} This work addresses a core challenge in enhancing personalization for \textbf{Web-scale recommender systems}: how to deeply and dynamically align the general world knowledge of Large Language Models (LLMs) with the heterogeneous behavioral patterns of individual users on \textbf{the Web}. Our research directly contributes to the 'User Modeling, Personalization, and Recommendation' track by proposing a novel framework to advance user modeling, with demonstrated effectiveness on large-scale \textbf{online platforms}.
% \end{quote}

\section{Introduction}
Recommender systems (RSs) as one of the solutions to alleviate the problem of information overload, are widely applied in various service platforms \cite{wang2021survey, wang2024rethinking, lin2024enhancing,wang2025home}. 
However, traditional RSs face significant limitations in addressing challenges such as cold-start and the long-tail distribution \cite{yang2024unifying, deng2025heterrec}.
In recent years, rapidly developing large language models (LLMs) have effectively alleviated these issues by leveraging world knowledge.
With the exponential growth of model scale and training corpora, LLMs have demonstrated the ability to implicitly encode world knowledge through self-supervised learning \cite{achiam2023gpt,bai2023qwen,liu2024deepseek,touvron2023llama}.
% Recent years have witnessed an exponential growth in both model scale and training corpora, enabling Large Language Models (LLMs) to implicitly encode world knowledge through self-supervised learning 
This knowledge encompasses factual information, commonsense logic, and abstract concepts, which are stored in distributed parameters \cite{yu2023kola, guu2020retrieval, petroni2019language}. 
% The parameterized representation allows LLMs to perform cross-domain reasoning and generalization capabilities based on contextual information. 
% Leveraging the generalization capabilities of LLMs, LLM-based RSs (LRSs) reduce dependence on behavior data \cite{boka2024survey}, and are emerging as a novel paradigm \cite{zhao2024recommender, liu2023chatgpt, wang2024rethinking}. 
Leveraging the reasoning and generalization abilities from the world knowledge, LLM-based RSs (LRSs) reduce dependence on behavioral data \cite{boka2024survey}, and are emerging as a novel paradigm \cite{zhao2024recommender, liu2023chatgpt, wang2024rethinking}. 

\begin{figure}[t]
  \centering
  \includegraphics[width=0.9\linewidth]{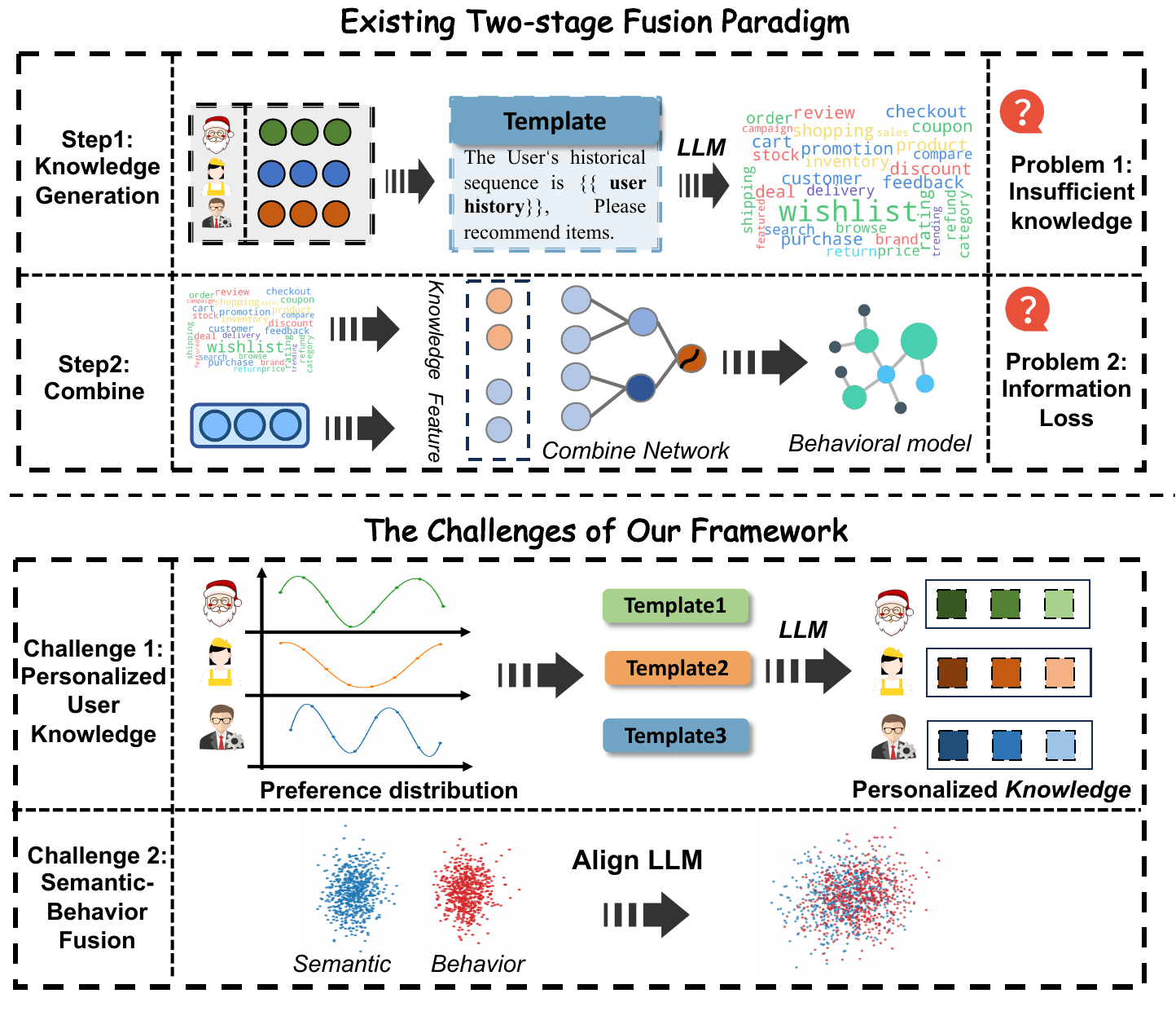}
  \caption{The problems of the existing paradigm and the challenges of our framework. }
  \label{fig1}
  \vspace{-0.5cm} %调整图片与上文的垂直距离
\end{figure}

Early LRSs employed LLMs as encoders, feeding generated embeddings as multimodal features into RSs \cite{morec,embedding-1,embedding-2,embedding-3,embedding-4}. However, this approach leverages LLMs solely for static content understanding, failing to exploit the predictive capabilities of  LLMs' world knowledge. Instead, it risks introducing redundant information such as image backgrounds, which could degrade recommendation performance \cite{wu2024survey}. To harness the world knowledge and reasoning capabilities of LLMs, recent studies have proposed a prompt-based two-stage knowledge fusion paradigm \cite{R4, KAR, lin2024clickprompt}. Specifically, in the first stage, user profiles, historical behavior sequences, and multimodal item features are integrated into a structured prompt template to guide LLMs to effectively leverage their pre-trained knowledge for generating semantically enriched implicit knowledge representations. In the second stage, deep neural networks or attention mechanisms are employed to perform fusion between the semantic knowledge generated by LLMs and behavioral features. However, as shown in Figure~\ref{fig1}, existing works suffer from certain limitations. \textbf{Firstly}, they predominantly adopt fixed structured prompt templates, which struggle to adapt to the multi-dimensional heterogeneity of user interests \cite{R4}. For instance, modeling behavioral sequences for users A and B respectively emphasizes demographic features (\textit{e.g.}, gender preferences) versus age-related life cycle characteristics.
When LLMs cannot dynamically adjust semantic focus through prompts, the generated implicit knowledge representations fail to capture the underlying interest distributions. 
\textbf{Secondly}, these structured prompt templates inherently contain numerous ID features (\textit{e.g.}, user and item IDs) that are semantically opaque to LLMs, which lead to significantly reduced accuracy. \textbf{Thirdly}, two-stage architectures treat LLMs as static knowledge sources achieving only superficial alignment between semantic knowledge and behavioral features at the output stage, while failing to model deep consistency across semantic and behavioral spaces \cite{deng2025csmf, xing2025esans}. To address these limitations, as shown in Figure~\ref{fig1}, we propose two research questions: \textbf{(i) How can LLMs generate personalized semantic knowledge that is strongly aligned with user behavioral patterns? \textbf{(ii)} How to establish effective alignment between the semantic space of LLMs and the behavioral space in RSs?}

% To address these challenges, we first investigate the variations in semantic knowledge generated by LLMs under different prompting strategies.
To address these challenges, we first analyze how semantic representations elicited from LLMs vary across different prompting strategies. Specifically, we construct 4 structured prompt templates with distinct emphasis patterns, and conduct experiments on diverse user cohorts within real-world e-commerce scenarios, leading to the following key findings:\textbf{(i)} The alignment between prompts and user characteristics significantly influences the value of semantic knowledge. Customized prompts tailored to specific user groups enable the semantic knowledge to achieve greater improvements in recommendation accuracy.\textbf{ (ii)} Not all semantic knowledge generated by LLMs proves beneficial to RSs. 
% Due to their probabilistic text generation mechanism, the inherent randomness in LLM outputs introduces noise into downstream recommendation tasks. 
Due to the probabilistic nature of LLM text generation, the inherent randomness in their outputs introduces noise into downstream recommendation tasks. Building on these findings, we propose an end-to-end LLMs and RSs \textbf{Co}llaboration-\textbf{Co}ntradiction fusion framework (\textbf{CoCo}), which consists of two stages: (1) \textbf{Collaboration enhancement}: Personalized semantic knowledge generation and (2) \textbf{Contradiction elimination}: behavior-semantic space alignment. 
% In the collaboration phase, we construct a structured codebook by leveraging the discrete encoding paradigm of VQ-VAE, where semantically initialized soft prompt candidates are organized as codebook entries. Subsequently, we perform similarity matching between user profile feature vectors and the codebook's soft prompt embeddings to dynamically select the top-m most relevant prompts as private prompts for each user. Concurrently, we designate a set of cross-domain universal prompts as supplementary components to ensure basic semantic knowledge generation when user information is insufficient. After obtaining the mixed semantic outputs containing multiple prompt responses from LLMs, we employ a cross attention mechanism to extract distinct semantic knowledge corresponding to each individual prompt.
In the collaboration enhancement phase, we leverage the discrete encoding paradigm of Vector Quantization (VQ) \cite{van2017neural,kingma2013auto,lee2022autoregressive,dong2020hawq} to dynamically generate $m$ soft prompts that maximize performance gains for each user. The cross attention mechanism \cite{vaswani2017attention} is then applied to extract the corresponding semantic knowledge from the mixed semantic outputs of LLMs for each individual prompt.
% In the contradiction phase, we dynamically assess the validity of LLM outputs at the user-level granularity. 
In the contradiction elimination phase, we dynamically evaluate at the user-level granularity whether LLMs' outputs provide a benefit to the recommendation system.
% For users associated with invalid knowledge,
When semantic knowledge fails to improve the performance of RSs, we employ the LoRA fine-tuning mechanism \cite{hu2022lora} to adjust LLM parameters, progressively aligning the semantic space of LLMs more closely with the behavioral space derived from user interaction sequences. 
% The CoCo framework synergizes collaboration and contradiction mechanisms by generating personalized semantic knowledge through dynamic prompts to enhance recommendation systems, while achieving semantic-behavioral space alignment via LoRA parameter optimization.

We conduct extensive assessments of CoCo across 2 public datasets and 1 industry-scale real-world dataset, comparing its performance against 7 state-of-the-art (SOTA) baselines (\textit{e.g.}, KAR \cite{KAR}, $R^4ec$ \cite{R4}) across 5 backbones. CoCo consistently achieves the best performance in all comparison scenarios, demonstrating up to 8.58\% improvement over the base models in recommendation effectiveness. Furthermore, we have deployed CoCo on a large-scale commercial advertising platform, where online A/B testing revealed a 1.91\% increase in advertising revenue and 0.64\% growth in gross merchandise volume (GMV). These empirical results demonstrate the significant practical implications of CoCo for real-world RSs.

Our contributions can be summarized as follows:

\begin{itemize}[noitemsep, topsep=0pt, leftmargin=*]
\item We systematically analyze the gains of semantic knowledge generated by LLMs, demonstrating that the alignment between prompts and user features significantly impacts the value of knowledge, and not all LLM outputs are beneficial to RSs.
\item We propose an end-to-end fusion framework named CoCo, which achieves semantic and behavioral space alignment through collaboration enhancement and contradiction elimination. 
\item Our proposed framework can be integrated into any existing RS. Extensive experiments on multiple public datasets and a large-scale industrial dataset demonstrate that CoCo surpasses 7 baselines across all evaluation metrics.
\end{itemize}

\section{Related Work}
% In this section, we first introduce the early world knowledge-enhanced LRSs, followed by the latest prompt-based two-stage fusion paradigm in this field.
\subsection{LRSs Enhanced by World Knowledge}
Early research in LLM-based recommendation aimed to leverage LLMs as a new backbone for RSs \cite{zhang2023prompt,wu2021userbert,penha2020does,sileo2022zero,sun2023chatgpt}, similar to their success in natural language processing and computer vision. A representative example is P5 \cite{p5}, which formulates a unified framework encompassing a variety of recommendation tasks, including sequential recommendation, rating prediction, and explanation generation, by transforming user behavior data into natural language sequences and feeding them into a pretrained language model to execute recommendation tasks.
However, a significant gap between the LLMs’ pretraining objectives and downstream recommendation tasks often leads to suboptimal performance. To address this mismatch, TALLRec \cite{tallrec} adopts instruction-tuning to make lightweight adaptations to LLMs. Nevertheless, as model size increases, the latency and computational cost of generating recommendation results via LLMs grow considerably, posing challenges for deployment in real-time applications.
Another line of work utilizes LLMs as auxiliary modules to enrich traditional recommendation pipelines. In this paradigm, LLMs encode multimodal features such as text and images, injecting world knowledge into conventional models through content-aware representations \cite{li2023exploring,li2023ctrl}. For instance, UniSRec \cite{unisrec} encodes item-related textual descriptions using a BERT-based \cite{koroteev2021bert} model, enabling the learning of transferable item embeddings across different domains. MoRec \cite{morec} integrates pretrained text and image encoders to capture rich content information, enhancing the expressiveness of the representations.

However, this paradigm treats LLMs primarily as feature encoders and does not fully explore the full potential of world knowledge reasoning embedded within LLMs. Moreover, multimodal inputs often contain noise irrelevant to user behavior, potentially affecting performance if directly concatenated with recommendation features. Therefore, recent research has focused on how to better enable RSs to utilize the world knowledge of LLMs.

\subsection{Prompt-Based Two-Stage Fusion in LRSs}
To effectively utilize the reasoning capabilities of LLMs grounded in world knowledge, recent works have introduced a prompt-based two-stage knowledge fusion paradigm. 
For example, KAR \cite{KAR} constructs user-side preference prompts and item-side factual prompts, feeding them into LLMs to obtain personalized reasoning and factual knowledge. These two types of knowledge are then passed through a knowledge adaptation module based on a mixture-of-experts (MoE) \cite{jacobs1991adaptive} framework, which compresses the outputs into low-dimensional representations compatible with recommendation features before integration into the recommendation model.  % 怎么介绍这一个方法写这么长
Similarly, $R^4$ec \cite{R4} proposes a knowledge reflection mechanism that iteratively evaluates the utility of retrieved knowledge and updates the knowledge before incorporating it into the model.

Existing works in this area are typically two-stage and rely on static prompt templates, which constrain the diversity of semantic knowledge generated by LLMs.
Furthermore, the fusion between RSs and LLMs is conducted at the feature dimension, without fundamentally aligning the behavioral and semantic spaces.
% In contrast to prior work, our approach addresses these limitations by enabling the generation of diverse semantic knowledge personalized to user-specific features and achieving a deep fusion between the semantic and behavioral spaces. 
In contrast to prior work, our framework generates diverse user-specific semantic knowledge and deeply fuses semantic and behavioral spaces.

\section{Pilot Experiments}
\subsection{Preliminaries}
% In this section, we formally define the retrieval task in recommendation systems. 
In this section, we formulate the recommendation task for RSs.
Given a user's sequential interaction sequence $S_{u} = \{i_{1}, i_{2}, ..., i_{T} \}$ from the item set $I$, the goal is to predict the next item $i_{T+1} \in I$. Each item $i$ is described by a feature set $d_i$. The model learns to map the user sequence $S_{u}$ and item features $d_i$ into representations $\textbf{u}$ and $\textbf{v}_{i}$, respectively. Following prior work \cite{sasrec, cobra}, we optimize the model using the InfoNCE contrastive loss \cite{mikolov2013distributed}. This objective trains the model to pull the user representation $\textbf{u}$ closer to the ground-truth (positive) item representation $\textbf{v}^{+}$ while pushing it away from $N$ negative item representations $\{ \mathbf{v}_j^{-} \}^{N}_{j=1}$:
\begin{equation}
\mathcal{ L}_{r}=- log \frac {exp(\mathbf{u}^T \mathbf{v}^+ / \tau)} {exp(\mathbf{u}^T \mathbf{v}^{+} / \tau) + \sum_{j=1}^{N} exp(\mathbf{u}^T \mathbf{v}_{j}^{-} / \tau)},
\label{eq:ams}
\end{equation}
where \(\tau\) is a temperature hyperparameter, and the negatives are sampled from other items in the mini-batch \cite{zheng2022multi,xing2025esans}.

\subsection{The Effectiveness of Semantic Knowledge}
\begin{figure}[ht]
  \includegraphics[width=0.48\textwidth]{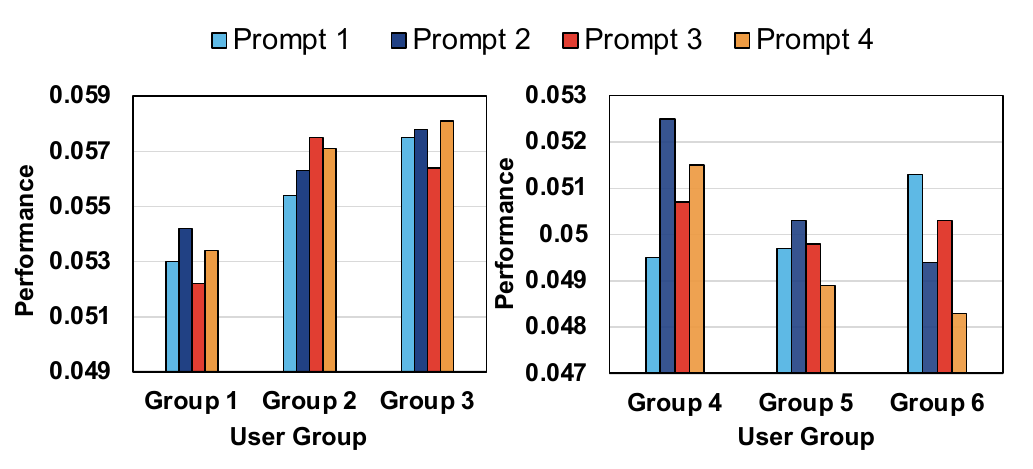}
  \caption{The performance comparison across 4 distinct prompts: Prompt 1 denotes the universal prompt, Prompt 2 is guided by age-based features, Prompt 3 is guided by gender-based features, and Prompt 4 is guided by item category.}
  \label{fig:prompt_data}
\end{figure}

To investigate the variations in semantic knowledge gains from LLMs under different prompt guidance, we conduct a pilot experiment in real-world e-commerce scenarios. Specifically, we randomly sample 6 non-overlapping user groups of 10,000 individuals each. Subsequently, we construct 4 distinct prompt templates: structured universal, age-guided, gender-guided, and item category-guided prompts, with detailed implementations provided in Appendix \ref{appd:pilot}. We then heuristically incorporate the semantic knowledge as features into a transformer-based RS \cite{pancha2022pinnerformer}. Each experiment is repeated 10 times with averaged results, and the performance is evaluated using Recall@5 \cite{tiger, unisrec}. As illustrated in Figure~\ref{fig:prompt_data}, we observe distinct performance gains across different prompts. Certain user groups exhibit significant improvements when employing feature-guided prompts: for instance, the gender-guided prompt achieves the highest performance gain for Group 2 users, while the age-guided prompt demonstrates optimal results for Group 4 users. This phenomenon indicates that the effectiveness of prompt-guided semantic knowledge is nonuniformly distributed across different users, necessitating dynamic optimization strategies tailored to the distinct contexts of users.
% This phenomenon indicates that the adaptability of prompt-guided semantic knowledge to specific user characteristics follows a non-uniform distribution, necessitating dynamic optimization strategies tailored to the distinct contexts of user.

\subsection{The Magnitude of Knowledge Gains in RSs}
\begin{figure}[ht]
\centering
\begin{minipage}[b]{0.48\columnwidth}
\centering
\includegraphics[width=\linewidth]{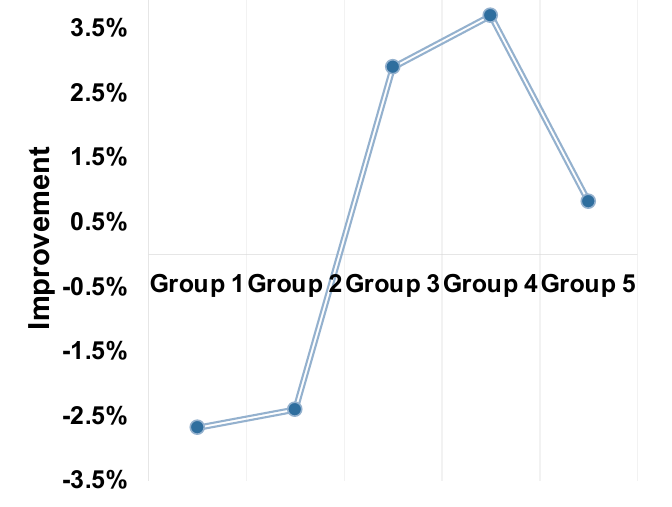}
\end{minipage}\hfill
\begin{minipage}[b]{0.48\columnwidth}
\centering
\includegraphics[width=\linewidth]{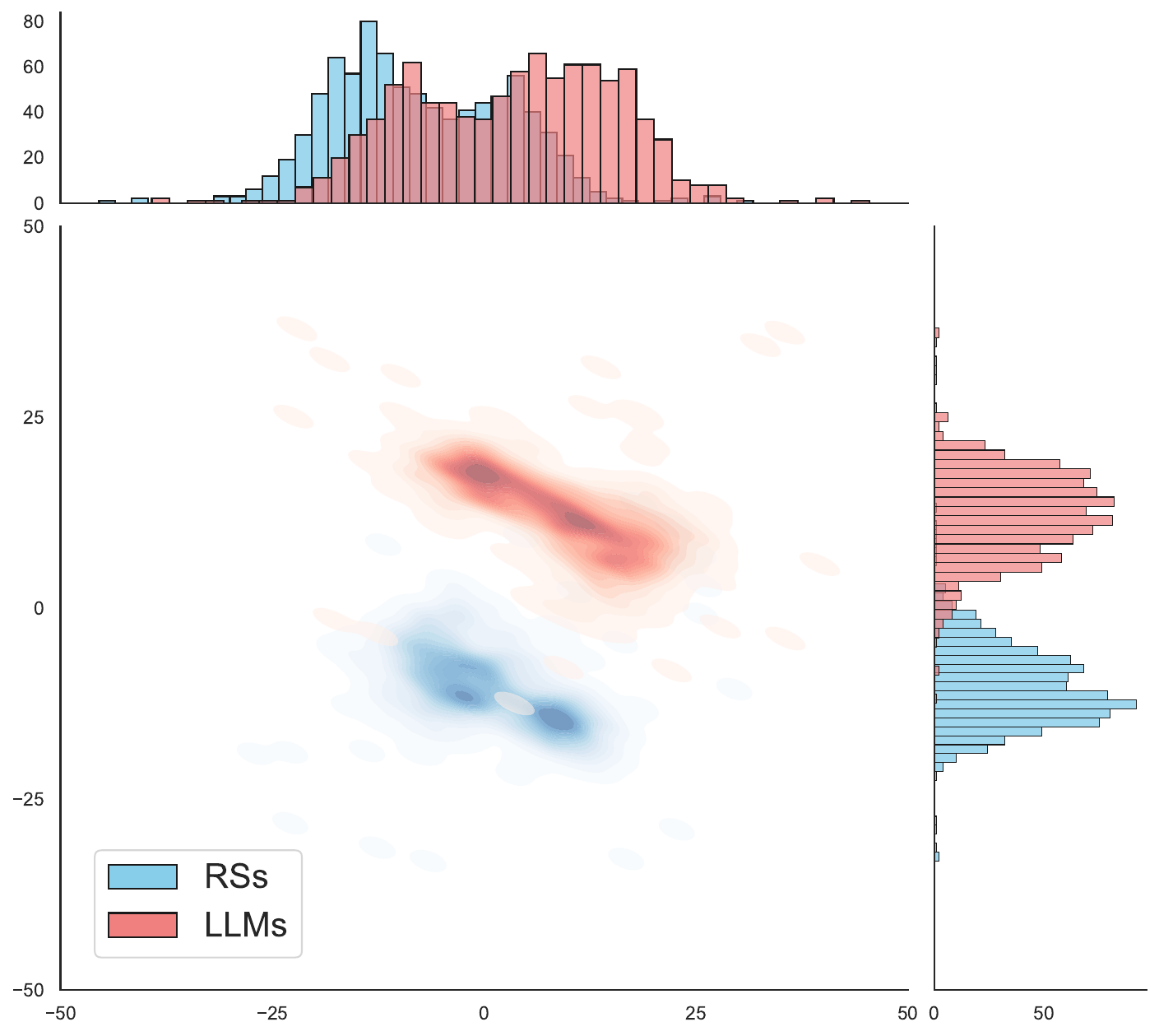}
\end{minipage}
\caption{(a) Performance gains of RSs from LLM-generated knowledge across 5 user groups (left). (b) Visualization of the semantic space represented by LLMs and the behavioral space represented by RSs (right).}
\label{fig:two-minipage}
\end{figure}

We further investigate the magnitude of gains that semantic knowledge provides to RSs. 
Specifically, we conduct experiments on 5 user groups (10,000 each), comparing a baseline model without LLM integration to an enhanced model incorporating knowledge from standard prompts.
% Specifically, we randomly select 5 groups of 10,000 users and conduct experiments comparing knowledge derived from standard prompts with the baseline without LLM involvement. 
The results are presented in Figure~\ref{fig:two-minipage} (left). Our findings reveal that semantic knowledge does not consistently improve recommendation performance and in some groups the system achieves better results without semantic knowledge input. As visualized in Figure~\ref{fig:two-minipage} (right) through t-SNE \cite{t-SNE}, there exists significant distributional divergence between LLM-generated semantic vectors and user behavior vectors. This discrepancy arises fundamentally from the distinct training objectives of LLMs and RSs: while LLMs employ self-supervised learning on massive textual corpora, RSs model user-item interaction sequences. The inherent distributional mismatch in latent space, compounded by the stochastic nature of LLM decoding processes, leads to negative knowledge injection effects when integrating these representations into RSs.

\begin{figure*}[ht]
  \centering
  \includegraphics[width=1.0\textwidth]{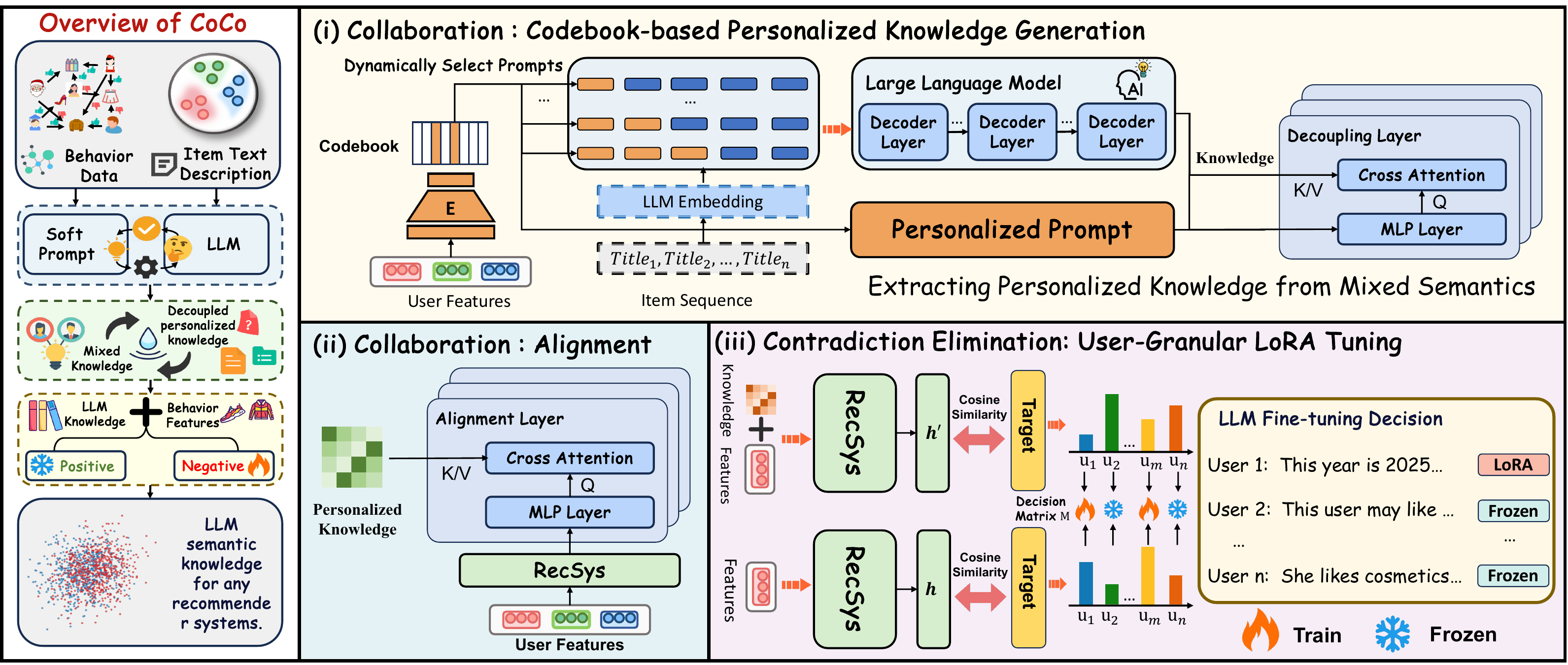 } % Reduce the figure size so that it is slightly narrower than the column.
  \caption{The overview of CoCo: An end-to-end collaboration-contradiction fusion framework integrating large language models and recommendation systems. 
  In the collaboration enhancement phase, we dynamically generate semantic knowledge for each user that is highly aligned with their behavioral patterns. In the contradiction elimination phase, we achieve alignment between the behavioral space and the semantic space through user-level LoRA fine-tuning.
  % In the collaboration phase, we dynamically generate user-specific semantic knowledge aligned with behavioral patterns. In the contradiction elimination phase, we align behavioral and semantic spaces via user-level LoRA fine-tuning.
  }
  % In the collaboration enhancement phase, we dynamically generate semantic knowledge for each user that is highly aligned with their behavioral patterns. In the contradiction elimination phase, we achieve alignment between the behavioral space and semantic space through user-level LoRA fine-tuning.}
  \label{fig:4}
\end{figure*}
\section{Methodology}
\subsection{Overview of CoCo}
Drawing insights from our pilot experiments, we propose CoCo, an end-to-end collaboration-contradiction fusion framework that integrates LLMs and RSs, as illustrated in Figure~\ref{fig:4}. 
CoCo aims to enhance RSs with LLMs from two perspectives: (1) Collaboration enhancement: dynamically generating personalized semantic knowledge to support the recommendation; and (2) Contradiction elimination: fine-tuning the LLM when semantic knowledge produces negative effects to align the semantic and behavioral spaces.

\subsection{Collaboration Enhancement}
\subsubsection{Personalized Prompt Generation based on Vector Quantization}
Inspired by our pilot experiments, we observe that a single static prompt cannot simultaneously maximize benefits for all users. An intuitive idea is to assign multiple tailored prompts to each user. However, this approach is impractical in real-world scenarios due to two main challenges. First, processing numerous prompts in parallel for each user would significantly increase the LLM's inference cost, leading to unacceptable latency. Second, concatenating too many prompts can introduce semantic conflicts and degrade the quality of the generated knowledge \cite{liu2023lost}. To address these issues, our goal is to build a large candidate prompt pool and dynamically select a small, optimal subset of $m$ prompts for each user from a total of $K$ candidates, where $m$ varies across users and $m \ll K$.
% Inspired by pilot experiments, we observe that a single static prompt cannot simultaneously maximize benefits for all users. An intuitive idea is to assign multiple tailored prompts for each user. However, this approach is impractical in real-world scenarios. First, processing numerous prompts in parallel for each user would significantly increase the LLM's inference cost, leading to unacceptable latency for real-time recommendation scenarios. Second, the risk of semantic conflicts rises, and an overly long prompt can degrade the quality of generated semantic knowledge \cite{liu2023lost}.
% To address these issues, we propose a personalized prompt generation method inspired by VQ-VAE, which generates individualized prompts by quantizing user representations into a discrete, learnable codebook. Specifically, we maintain a candidate semantic prompt pool $\mathcal{Z}=\{\mathbf{z}_1, \mathbf{z}_2, \dots, \mathbf{z}_K\}$ comprising $K$ prompts, from which we select the $m$ most beneficial ones for each user, where $m$ varies for different users and $m \ll K$.

To achieve this, we first need to construct the candidate prompt pool. 
While pilot experiments demonstrated the effectiveness of manually selected semantic prompts in specific scenarios, it is nearly impossible to enumerate all potentially effective prompts, especially considering the millions of users in RSs. 
Furthermore, structured templates often contain raw IDs, which lack semantic richness and may introduce noise.
To overcome these challenges, we propose representing the candidate pool as a set of soft prompts, trainable continuous vectors, to replace manually crafted semantic templates. We define this candidate pool as $\mathcal{Z}=\{\mathbf{z}_1, \mathbf{z}_2, \dots, \mathbf{z}_K\}$, where each $\mathbf{z}_i \in \mathbb{R}^d$ is a learnable soft prompt embedding. These soft prompts are jointly trained with the RS, allowing them to self-adapt and compress personalized semantic information.
% Through joint training with RS, soft prompts can automatically discover implicit semantic correlations that contribute to better recommendation outcomes. Moreover, soft prompts feature a self-adaptive mechanism that continuously compresses and integrates personalized semantic prompts during training, thus reducing the scale of $K$.
% To overcome these challenges, we introduce soft prompts, a set of trainable continuous vector parameters, to replace manually crafted semantic templates. Through joint training with RS, soft prompts can automatically discover implicit semantic correlations that contribute to better recommendation outcomes. Moreover, soft prompts feature a self-adaptive mechanism that continuously compresses and integrates personalized semantic prompts during training, thus reducing the scale of $K$.
% Therefore, to overcome these challenges, we propose to represent the candidate pool as a set of soft prompts—trainable continuous vectors that can automatically discover implicit semantic correlations beneficial for recommendation. We define this candidate pool, which we term the prompt codebook, as 

With the prompt codebook $\mathcal{Z}$ established, our next goal is to select the $m$ most relevant prompts for each user, where $m$ is dynamically determined based on individual user characteristics. Inspired by the effectiveness of Vector Quantization (VQ) in modeling discrete latent spaces \cite{van2017neural,lee2022autoregressive}, we adopt a VQ-based framework for this selection task. The $K$ soft prompt vectors in $\mathcal{Z}$ naturally serve as the discrete codebook. To ensure stability and accelerate convergence, we initialize these soft prompts using universal semantic prompt vectors encoded by an LLM, rather than with random initialization. For each user $u$, we first encode their features (\textit{e.g.}, user ID and interaction sequence) into a unified representation $\mathbf{e}_{u}$. We then compute the cosine similarity between $\mathbf{e}_{u}$ and each codeword ${\mathbf{z_{i}}}$ in the codebook. Finally, these similarity scores are normalized into a probability distribution over the prompts using the softmax function \cite{krizhevsky2012imagenet}:
% After building a pool of soft prompt candidates, our goal is to select $m$ most relevant prompts for each user, where $m$ is dynamically determined based on individual user characteristics. Given the advantages of Vector Quantization (VQ) in modeling discrete latent spaces and efficient encoding, we incorporate a VQ-VAE based framework. Specifically, the $K$ soft prompt vectors in the candidate pool are organized into a learnable codebook, serving as discrete semantic anchors for personalized prompt selection. To ensure stability and accelerate convergence, we initialize these soft prompts with universal semantic prompt vectors encoded by an LLM, rather than with random initialization. Then we concatenate the encoded user features (\textit{e.g.}, user ID and interaction sequence) into a unified feature vector $\mathbf{e}_{u}$ and compute the cosine similarity between $\mathbf{e}_{u}$ and each code word ${\mathbf{z_{i}}}$. Finally, the similarity scores are normalized into the [0,1] range via softmax, yielding a probability distribution over the most relevant prompts \cite{krizhevsky2012imagenet}:
\begin{equation} \label{equ:softmax}
s_{i} = \operatorname{Softmax}\left(\operatorname{cos}(\mathbf{e}_{u}, \mathbf{z}_{i})\right), \quad i=1,2,\dots,K.
% p_{i} = \frac{\exp(s_{i})}{\sum_{j=1}^{K} \exp(s_{j})}
\end{equation}

Based on the similarities, we set a threshold $\theta$ and select prompts with $s_i > \theta$ as the input prefix to the LLM for each user. To ensure the model can generate foundational semantic knowledge even with insufficient user information, we concatenate a shared structured semantic prompt vector $\mathbf{z}_{share}$ before each user's private prompt. 
% Then, each selected soft prompt generates a corresponding learnable lookup embedding $\mathbf{p}_i$, which is derived from the codeword ${i}$ entry via the quantization process. Finally, all selected prompt vectors are concatenated with the LLM-encoded user profile vectors $\mathbf{h}_{text}$ as the input $\mathbf{h}_l$ of the LLM decoding layer, where $\mathbf{h}_{text}$ is the text feature embedding of the user's historical interaction items encoded by LLM:
% TODO:这个应该还要再改一下，暂时放在这
% \begin{equation}\label{equ:concentrate}
% \left[h_{i, l+1, z}\right]_{z=1}^Z=\left[p_{i, l, k}\right]_{k=1}^K \oplus\left[h_{i, l, z}\right]_{z=1}^Z ,
% \end{equation}

The selected soft prompts $\{\mathbf{z}_{i} | s_{i}>\theta\}$, which now serve as the user's personalized prompts, are concatenated with the shared prompt $\mathbf{z}_{share}$ to form the final prompt matrix $\mathbf{h}_{prompt}$. This matrix is then concatenated with the user's textual profile embedding $\mathbf{h}_{text}$ to create the final input $\mathbf{h}_{l}$ for the LLM's decoding layers. Here, $\mathbf{h}_{text}$ is obtained by encoding the titles of the user's historical interaction items using the same LLM. The complete input is formulated as:
\begin{equation}\label{equ:final_input_combined}
\begin{aligned}
\mathbf{h}_{prompt} &= [\mathbf{z}_{shared}, \{\mathbf{z}_{i} | s_{i}>\theta\}], \\
\mathbf{h}_l &= [\mathbf{h}_{prompt}, \mathbf{h}_{text}].
\end{aligned}
\end{equation}

Additionally, all user vectors are padded to a uniform length to satisfy the LLM's input requirements.
The training of the soft prompts and the codebook is guided by the VQ objective \cite{van2017neural}, which incorporates the quantization loss $\mathcal{L}_Q$:
% For training the soft prompts and optimizing the codebook, we adopt the VQ-VAE objective, including the quantization loss $\mathcal{L}_Q$:
\begin{equation}
\mathcal{L}_Q=\left\|\mathbf{e}_{u}-\mathbf{z}_k\right\|_2^2,
% \mathcal{L}_Q=\left\|\operatorname{sg}\left(\mathbf{z}_{user}\right)-\mathbf{z}_k\right\|_2^2,
%\quad\mathcal{L}_C=\left\|\mathbf{z}_{user}-\operatorname{sg}\left(\mathbf{z}_k\right)\right\|_2^2,
\end{equation}
where $\mathbf{z}_k$ represents the selected codeword. Consequently, the codebook is optimized to learn discrete user-specific representations.
% $\mathcal{L}_Q$ aims to minimize the mean squared error between the user features and nearest neighboring prompt codeword vector, updating the codebook to match the encoder output. 
% $\mathcal{L}_C$ constrains the user features from drifting away from the codeword vectors, which helps prevent the codebook from being updated too slowly.

\subsubsection{Semantic Decoupling based on Orthogonal Constraint}
Concatenating multiple prompts as input may result in LLM outputs that combine conflicting or inconsistent semantic information, introducing ambiguity or errors in the model's predictions.
Ideally, each prompt should be fed into the LLM independently to obtain pure semantic knowledge. However, this would increase inference costs, making it impractical for real-time RSs. 
Therefore, we decouple the mixed semantic output into knowledge fragments corresponding to each prompt. 
% Specifically, we use the cross attention to separate the pure semantic knowledge $\mathbf{h}_{pure}$ from the mixed output $\mathbf{h}_{mix}$, where the prompt embeddings form the query and the mixed semantic knowledge as the key and value:
Specifically, we employ the cross attention with the mixed semantic knowledge $\mathbf{h}_{mix}$ from the LLM as both Key and Value, and the final prompt matrix $\mathbf{h}_{prompt}$ as the Query:
% Specifically, we use the cross attention where the mixed semantic knowledge $\mathbf{h}_{mix}$ serves as the Key and Value. The Query is formed by prompt matrix $\mathbf{h}_{prompt} = [\mathbf{z}_{shared}, \mathbf{p}_1, \dots, \mathbf{p}_m]$, which concatenates a shared prompt embedding with $m$ personalized ones. The decoupling is then performed as:
% \begin{equation} \label{equ:decoupling}
%     \mathbf{h}_{pure} = \text{CrossAttention}(\mathbf{h}_{pure}W^{1}_{Q}, \mathbf{h}_{mix}W^{1}_{K}, \mathbf{h}_{mix}W^{1}_{V}),
% \end{equation}
\begin{equation}
% \begin{gathered}
\label{equ:decoupling_concat}
    % \mathbf{Q}_{prompt} = [\mathbf{z}_{shared}, \mathbf{p}_1, \dots, \mathbf{p}_m], \\
    \mathbf{h}_{pure} = \text{CrossAttention}(\mathbf{h}_{prompt} W^{1}_{Q}, \mathbf{h}_{mix}W^{1}_{K}, \mathbf{h}_{mix}W^{1}_{V}),
% \end{gathered}
\end{equation}
% \begin{equation} \label{equ:decoupling}
%     \mathbf{h}_{pure} = \text{CrossAttention}([\mathbf{z}_{shared}, \left[\mathbf{z}_{i}\right]_{i=1}^m]W^{1}_{Q}, \mathbf{h}_{mix}W^{1}_{K}, \mathbf{h}_{mix}W^{1}_{V}),
% \end{equation}
where $W^1_Q$, $W^1_K$, $W^1_V$ are learnable projection matrices. To prevent semantic redundancy within the prompt candidate pool, we further introduce an orthogonal constraint loss $\mathcal{L}_{ortho}$ to minimize the similarity between prompts and enhance diversity:
\begin{equation} \label{equ:ortho}
\mathcal{L}_{ortho}=\sum_{i=1}^K \sum_{j=j+1}^K \operatorname{cos} \left(\mathbf{z}_i, \mathbf{z}_j\right),
\end{equation}
where $\text{cos}(\mathbf{z}_i, \mathbf{z}_j)$ represents the cosine similarity between prompts $\mathbf{z}_i$ and $\mathbf{z}_j$. $\mathcal{L}_{ortho}$ ensures that each prompt is semantically complementary rather than redundant.

\subsubsection{Semantic-Behavioral Alignment}
After obtaining the decoupled semantic knowledge $h_{pure}$, we employ an attention-based module to align it with the behavioral representations $h_{RS}$ from the RSs. 
First, using $h_{RS}$ as the Query, the Key and Value are formed by concatenating $h_{RS}$ with $h_{pure}$ and then projecting the result through a linear layer.
% First, with the behavioral output $h_{RS}$ serves as the query, we concatenate behavioral features and semantic knowledge as the key and value to perform cross attention. 
This enables the RSs to selectively attend to the semantic features most relevant to the prediction task.
The formulation of cross attention is:
% \begin{equation} \label{equ:cross-attn}
% \begin{aligned}
% \mathbf{Q}, \mathbf{K}, \mathbf{V} & = h_{RS}W_Q, h_{concat}W_K, h_{concat}W_V,\\
% h_{aligned} & = \operatorname{CrossAttention}\left(\mathbf{Q}, \mathbf{K}, \mathbf{V}\right),
% \end{aligned}
% \end{equation}
\begin{equation}
    h_{aligned} = \operatorname{CrossAttention}\left(h_{RS}W^{2}_{Q}, h_{concat}W^{2}_{K}, h_{concat}W^{2}_{V}\right),
\end{equation}
% where $h_{concat} = MLP(h_{RS} \oplus h_{pure})$ represents the fused representation derived from concatenation. 
where $h_{concat} = MLP([h_{RS}, h_{pure}])$ represents the fused representation derived from concatenation. 
% is the concatenated vectors. % dim?
% To further enhance the representation quality, we apply the self attention to capture global dependencies and promote a deep fusion between semantic knowledge and behavioral features: 
% \begin{equation} \label{equ:self-attn}
% % \boldsymbol{o}_i=\operatorname{SelfAttn}\left(\boldsymbol{o}_i^{(l)} \oplus \boldsymbol{o}_i^{(2)} \oplus \ldots \oplus \boldsymbol{o}_j^{(L)}\right)
% % \mathbf{h_{fused}} = \text{SelfAttention}(\mathbf{Q}=\mathbf{h_{aligned}}, \mathbf{K}=\mathbf{h_{aligned}}, \mathbf{V}=\mathbf{h_{aligned}})
% h_{fused} = \text{SelfAttention}(h_{aligned}W^{3}_{Q}, h_{aligned}W^{3}_{K}, h_{aligned}W^{3}_{V}),
% \end{equation}
Finally, we replace the original output of the RSs with the output of the alignment module to compute the recommendation loss $\mathcal{L}_r$. 
% The overall training objective is a combination of all losses:
% \begin{equation}
% \mathcal{L} = \mathcal{L}_r + \mathcal{L}_C + \mathcal{L}_Q + \mathcal{L}_{ortho} .
% \end{equation}

\begin{table*}[htbp!]
\centering
\caption{Performance comparison of various retrieval backbones on the 3 datasets. Best results are in bold. "$Impr.$" shows the relative improvement (\%) over the base model.}
\renewcommand{\arraystretch}{1.1} % 增加行间距
\resizebox{0.95\linewidth}{!}{
    \begin{tabular}{cc|cccc|cccc|cccc}
    \toprule
    \multirow{2}{*}{\textbf{Backbone}} & \multirow{2}{*}{\textbf{Method}} & \multicolumn{4}{c|}{\textbf{Beauty}} & \multicolumn{4}{c|}{\textbf{Toys}} & \multicolumn{4}{c}{\textbf{Industrial Dataset}}\\
     & & $R@5$ & $Impr.$ & $N@5$ & $Impr.$ & $R@5$ & $Impr.$ & $N@5$ & $Impr.$ & $R@5$ & $Impr.$ & $N@5$ & $Impr.$ \\ 
    \midrule
    
    \multirow{4}{*}{BERT4Rec} 
    & base 
    & 0.0203 & — & 0.0124 & — & 0.0116 & — & 0.0069 & — & 0.0418 & — & 0.0282 & — \\
    & KAR  
    & 0.0206 & $+1.48\%$ & 0.0127 & $+2.42\%$ & 0.0117 & $+0.86\%$ & 0.0070 & $+1.45\%$ & 0.0423 & $+1.20\%$ & 0.0284 & $+0.71\%$ \\
    & $R^4$ec  
    & 0.0207 & $+1.97\%$ & 0.0129 & $+4.03\%$ & 0.0118 & $+1.72\%$ & 0.0071 & $+2.90\%$ & 0.0427 & $+2.15\%$ & 0.0285 & $+1.06\%$ \\
    & CoCo (Ours)
    & 0.0213 & $\textbf{+4.93\%}$ & 0.0134 & $\textbf{+8.06\%}$ & 0.0121 & $\textbf{+4.31\%}$ & 0.0073 & $\textbf{+5.80\%}$ & 0.0439 & $\textbf{+5.02\%}$ & 0.0291 & $\textbf{+3.19\%}$ \\
    
    \midrule
    
    \multirow{4}{*}{FDSA} 
    & base 
    & 0.0267 & — & 0.0163 & — & 0.0228 & — & 0.0140 & — & 0.0445 & — & 0.0303 & — \\
    & KAR  
    & 0.0272 & $+1.87\%$ & 0.0166 & $+1.84\%$ & 0.0230 & $+0.88\%$ & 0.0142 & $+1.43\%$ & 0.0452 & $+1.57\%$ & 0.0305 & $+0.66\%$ \\
    & $R^4$ec  
    & 0.0275 & $+3.00\%$ & 0.0167 & $+2.45\%$ & 0.0234 & $+2.63\%$ & 0.0143 & $+2.14\%$ & 0.0458 & $+2.92\%$ & 0.0308 & $+1.65\%$ \\
    & CoCo (Ours)  
    & 0.0280 & $\textbf{+4.87\%}$ & 0.0170 & $\textbf{+4.29\%}$ & 0.0240 & $\textbf{+5.26\%}$ & 0.0145 & $\textbf{+3.57\%}$ & 0.0477 & $\textbf{+7.19\%}$ & 0.0315 & $\textbf{+3.96\%}$ \\
    
    \midrule
    
    \multirow{4}{*}{$S^3$Rec} 
    & base 
    & 0.0387 & — & 0.0244 & — & 0.0443 & — & 0.0294 & — & 0.0569 & — & 0.0476 & — \\
    & KAR  
    & 0.0388 & $+0.26\%$ & 0.0246 & $+0.82\%$ & 0.0454 & $+2.48\%$ & 0.0298 & $+1.36\%$ & 0.0576 & $+1.23\%$ & 0.0480 & $+0.84\%$ \\
    & $R^4$ec  
    & 0.0389 & $+0.52\%$ & 0.0247 & $+1.23\%$ & 0.0453 & $+2.26\%$ & 0.0296 & $+0.68\%$ & 0.0578 & $+1.58\%$ & 0.0482 & $+1.26\%$ \\
    & CoCo (Ours)  
    & 0.0398 & $\textbf{+2.84\%}$ & 0.0253 & $\textbf{+3.69\%}$ & 0.0457 & $\textbf{+3.16\%}$ & 0.0299 & $\textbf{+1.70\%}$ & 0.0602 & $\textbf{+5.80\%}$ & 0.0488 & $\textbf{+2.52\%}$ \\
    
    \midrule
    
    \multirow{4}{*}{SASRec} 
    & base 
    & 0.0387 & — & 0.0249 & — & 0.0463 & — & 0.0306 & — & 0.0493 & — & 0.0339 & — \\
    & KAR  
    & 0.0395 & $+2.07\%$ & 0.0255 & $+2.41\%$ & 0.0474 & $+2.38\%$ & 0.0309 & $+0.98\%$ & 0.0507 & $+2.84\%$ & 0.0345 & $+1.77\%$ \\
    & $R^4$ec  
    & 0.0402 & $+3.88\%$ & 0.0258 & $+3.61\%$ & 0.0476 & $+2.81\%$ & 0.0310 & $+1.31\%$ & 0.0509 & $+3.25\%$ & 0.0350 & $+3.24\%$ \\
    & CoCo (Ours)  
    & 0.0414 & $\textbf{+6.98\%}$ & 0.0265 & $\textbf{+6.43\%}$ & 0.0484 & $\textbf{+4.54\%}$ & 0.0315 & $\textbf{+2.94\%}$ & 0.0521 & $\textbf{+5.68\%}$ & 0.0355 & $\textbf{+4.72\%}$ \\
    
    \midrule
    
    \multirow{4}{*}{PinnerFormer} 
    & base 
    & 0.0516 & — & 0.0373 & — & 0.0585 & — & 0.0455 & — & 0.0697 & — & 0.0469 & — \\
    & KAR  
    & 0.0522 & $+1.16\%$ & 0.0377 & $+1.07\%$ & 0.0595 & $+1.71\%$ & 0.0458 & $+0.66\%$ & 0.0708 & $+1.58\%$ & 0.0473 & $+0.85\%$ \\
    & $R^4$ec 
    & 0.0536 & $+3.88\%$ & 0.0383 & $+2.68\%$ & 0.0597 & $+2.05\%$ & 0.0460 & $+1.10\%$ & 0.0711 & $+2.01\%$ & 0.0475 & $+1.28\%$ \\
    & CoCo (Ours)  
    & 0.0549 & $\textbf{+6.40\%}$ & 0.0405 & $\textbf{+8.58\%}$ & 0.0634 & $\textbf{+8.38\%}$ & 0.0476 & $\textbf{+4.62\%}$ & 0.0752 & $\textbf{+7.89\%}$ & 0.0497 & $\textbf{+5.97\%}$ \\
    
    \bottomrule
    \end{tabular}
}
\label{table:backbone_experiment_results}
\end{table*}
\subsection{Contradiction Elimination}
% In existing studies, the RSs often unconditionally accept semantic knowledge generated by LLMs. However, our pilot experiments reveal that LLM outputs are not always beneficial. 
In existing studies, RSs typically accept semantic knowledge generated by LLMs without validation. However, our pilot experiments reveal that LLM outputs are not always beneficial. In some cases, they may even degrade recommendation performance. A significant discrepancy often exists between the behavioral space, constituted by user interaction features like clicks and views, and the semantic space, formed by LLM-generated semantic knowledge embeddings. This divergence can result in LLM outputs that fail to reflect genuine user preferences. 

To address this, we propose a dynamic contradiction elimination fine-tuning strategy that assesses the utility of LLM outputs in real-time. When the LLM-generated knowledge is deemed ineffective, we selectively fine-tune the LLM to gradually narrow the gap between the semantic and behavioral spaces, ensuring that LLM-generated knowledge becomes increasingly aligned with user behavior over time.
% In existing studies, RSs typically accept semantic knowledge generated by LLMs without validation, assuming it naturally aligns with user preferences. However, our pilot experiments reveal that LLM outputs are not always beneficial. In some cases, they may even degrade recommendation performance.
% A key reason lies in the growing discrepancy between two distinct spaces: the behavioral space, defined by real user interactions such as clicks and views, and the semantic space, constructed from high-level concepts encoded by LLMs. Since LLMs are often pre-trained on general corpora rather than personalized behavior data, their generated knowledge may reflect plausible but irrelevant semantics, leading to a misalignment with actual user needs.
% To bridge this gap, we propose a dynamic contradiction elimination fine-tuning strategy that continuously evaluates the utility of LLM-generated outputs during training. When the model detects that certain semantic knowledge fails to contribute to accurate recommendations, it triggers selective fine-tuning on the LLM component, suppressing misleading signals and reinforcing behaviorally consistent semantics. This process enables the system to progressively narrow the divergence between the semantic and behavioral spaces, ensuring that LLM-generated knowledge becomes increasingly aligned with user behavior over time.
We first assess whether LLMs contribute positively to recommendation tasks. We compare the target item's embedding $\textbf{v}_t$ with two outputs: (1) the original RSs' output $h_{RS}$, and (2) the output after aligning with LLM-derived semantic knowledge $\textbf{h}_{aligned}$.
The indicator function and decision matrix $M$ can be formulated as:
\begin{equation} \label{equ:indicator}
M=\mathbb{I}((\text{cos}(h_{aligned}, \textbf{v}_{t}) > \text{cos}(h_{RS}, \textbf{v}_t))),
% M=\begin{cases}1 & \text { if } \text{cos}(h_{with\_LLM}, i_t) - \text{cos}(h_{without\_LLM}, i_t)>\theta_2, \\ 0 & \text { otherwise. } \end{cases}
\end{equation}
% 判定函数
where $\text{cos}$ denotes cosine similarity, and when the condition in the indicator function $\mathbb{I}(\cdot)$ holds, $\mathbb{I}=1$; otherwise, $\mathbb{I}=0$.
% An output is considered beneficial if it improves the similarity beyond a threshold $\theta_2$.
Based on this decision, we obtain a binary matrix $M \in \{0,1\}$, where a value of 1 indicates that the LLM output is beneficial. 
We then employ gradient masking to control the fine-tuning process, ensuring that the gradient update is performed only when the LLM output provides no benefit to the recommendation ($M=0$), leaving the effective parts frozen and unchanged.
The formal process is:
\begin{equation}
% s^{\prime}=\sqrt{g}(D) m \theta S+(1-m) \theta S
h_{aligned}^{\prime} = \text{sg}(M \odot h_{aligned}) + (1-M) \odot h_{aligned},
\end{equation}
where $\text{sg}(\cdot)$ denotes the stop-gradient operation and $\odot$ denotes element-wise multiplication.
% Furthermore, we adopt Low-Rank Adaptation (LoRA) to perform this fine-tuning in a parameter-efficient manner, by superimposing low-rank matrices onto the original LLM parameter matrices:
Furthermore, given LoRA's significant advantages in both parameter efficiency and task adaptability, we adopt LoRA for fine-tuning the LLM:
\begin{equation} \label{equ:lora}
W^{\prime}=W+A B, A \in \mathbb{R}^{d \times r}, B \in \mathbb{R}^{r \times d}, r \ll d,
\end{equation}
% \begin{equation}
% s_i^{\prime}=\left(\boldsymbol{W}+\boldsymbol{A}_i \mathbf{B}_i\right) \cdot \boldsymbol{x}_i
% \end{equation}
where $W$ is the original weight matrix, $A$ and $B$ are low-rank matrices with rank $r$. Only a small number of parameters (the elements of $A$ and $B$) need to be adjusted to adapt the model.
This parameter-efficient update mechanism avoids direct modification of the LLM's original parameters, thereby largely preserving the general world knowledge acquired during its pre-training phase. 

Furthermore, to ensure semantic alignment between $h_{RS}$ and the item representation $\textbf{v}_t$, we introduce an auxiliary loss $\mathcal{L}_{aux}$, which mirrors the InfoNCE formulation of $\mathcal{L}_r$ (Eq.~\ref{eq:ams}) by replacing the user representation $\textbf{u}$ with $h_{RS}$. The final training objective is:
% We further introduce an auxiliary loss, $\mathcal{L}_{aux}$, to align $h_{RS}$ with the item representation $\textbf{v}_t$. It follows the InfoNCE formulation of $\mathcal{L}_r$ (Eq.~\ref{eq:ams}), substituting $\textbf{u}$ with $h_{RS}$. The overall objective is:
% to ensure that the $h_{RS}$ and $\textbf{v}_t$ are semantically aligned, we introduce an auxiliary loss $\mathcal{L}_{aux}$. This loss adopts the same formulation as the $\mathcal{L}_r$ (Eq.~\ref{eq:ams}), but replaces the $\textbf{u}$ with $h_{RS}$. Finally, the overall training objective combines all losses:
\begin{equation}
\mathcal{L} = \mathcal{L}_r + \alpha \mathcal{L}_{aux} + \beta \mathcal{L}_{ortho}  + \gamma \mathcal{L}_{Q},
\end{equation}
where $\alpha$, $\beta$, and $\gamma$ are loss weights. The complete algorithm for CoCo is detailed in Appendix \ref{appd:algorithm}.

\begin{table*}[htbp!]
\centering
\caption{Performance Comparison with SOTA baselines. Best results are in \textbf{bold}; the second-best are underlined. Numbers in parentheses denote the relative percentage improvement of our method over the strongest baseline.}
\resizebox{0.95\linewidth}{!}{
    \begin{tabular}{c|cc|cc|cc}
    \toprule
    \multirow{2}{*}{\textbf{Method}} & \multicolumn{2}{c|}{\textbf{Beauty}} & \multicolumn{2}{c|}{\textbf{Toys}} &  \multicolumn{2}{c}{\textbf{Industrial Dataset}}\\
     & $R@5$ & $N@5$ &  $R@5$ & $N@5$ &  $R@5$ & $N@5$  \\ 
    \midrule
    UniSRec & 0.0329 & 0.0248 & 0.0429 & 0.0292 & 0.0594 & 0.0385 \\
    TALLRec & 0.0403 & 0.0295 & 0.0498 & 0.0327 & 0.0655 & 0.0421 \\
    Recformer & 0.0439 & 0.0317 & 0.0512 & 0.0345 & 0.0640 & 0.0418 \\
    Tiger & 0.0454 & 0.0321 & 0.0521 & 0.0371 & 0.0677 & 0.0446 \\
    COBRA & \underline{0.0537} & \underline{0.0395} & \underline{0.0619} & \underline{0.0462} & \underline{0.0716} & \underline{0.0480} \\
    % \midrule
    PinnerFormer & 0.0516 & 0.0373 & 0.0585 & 0.0455 & 0.0697 & 0.0469 \\
    PinnerFormer + CoCo & \textbf{0.0549}(\textit{+2.23\%}) & \textbf{0.0405}(\textit{+2.53\%}) & \textbf{0.0634}(\textit{+2.42\%}) & \textbf{0.0476}(\textit{+3.03\%}) & \textbf{0.0752}(\textit{+5.03\%}) & \textbf{0.0497}(\textit{+3.54\%}) \\
    % \midrule
    % \textit{Impr.} & \textit{2.23\%} & \textit{2.53\%} & \textit{2.42\%} & \textit{3.03\%} & \textit{5.03\%} & \textit{3.58\%} \\
    \bottomrule
    \end{tabular}
}
\label{table:baseline_experiment_results}
\vspace{-8pt}
\end{table*}

\section{Experiments}
To comprehensively evaluate CoCo's performance, we conduct extensive experiments to address the following research questions:
\begin{itemize}[noitemsep, topsep=0pt, leftmargin=*]
    \item \textbf{RQ1}: Does CoCo outperform state-of-the-art (SOTA) two-stage knowledge fusion baselines? Does it demonstrate robust adaptability across different backbone architectures?
    \item \textbf{RQ2:} How does CoCo compare with other LRSs paradigms in terms of performance?
    \item \textbf{RQ3:} Can scaling up LLMs' size enhance CoCo's performance?
    \item \textbf{RQ4:} What impact do hyperparameters have on its effectiveness?
    \item \textbf{RQ5:} How do different knowledge fusion strategies affect CoCo's performance?
    \item \textbf{RQ6:} Can CoCo integrate the semantic and behavioral spaces?
    \item \textbf{RQ7:} What are the contributions of each module in CoCo?
    \item \textbf{RQ8:} Can CoCo deliver performance improvements in real-world online recommendation platforms?
\end{itemize}

\subsection{Experimental Setup}
\subsubsection{Datasets and Evaluation Metrics.}
We evaluate our approach on \textbf{3 datasets}: 2 public datasets and 1 in-house industrial dataset. The public datasets are two subsets, "Beauty" and "Toys \& Games", from the Amazon Product Reviews collection \cite{AmazonDataset}. The industrial dataset is an in-house collection of user interaction logs from a leading Southeast Asian e-commerce platform, comprising 18 million users and 23 million interactions over a seven-month period in 2025. 
% The massive scale and inherent complexity of this real-world dataset provide a more challenging and realistic testbed for our proposed method. 
We adopt two top-K ranking metrics: Recall@5 (R@5) and NDCG@5 (N@5) \cite{tiger, unisrec}. 
Dataset statistics and preprocessing details are provided in Appendix \ref{appd:dataset}.

\subsubsection{Backbone Models.}
Our method is model-agnostic and can accommodate various retrieval models as its backbone. To demonstrate this generality, we integrate CoCo with 5 representative models that span different architectures: \textbf{SASRec} \cite{sasrec}, which uses unidirectional self-attentive network for sequential modeling; \textbf{BERT4Rec} \cite{bert4rec}, which employs a bidirectional transformer to capture contextual user interests; \textbf{FDSA} \cite{fdsa}, which learns fine-grained feature patterns via self-attention; $\textbf{S}^3$\textbf{Rec} \cite{s3rec}, which leverages self-supervised learning to enhance representations; and \textbf{PinnerFormer} \cite{pancha2022pinnerformer}, which models long-term behavior using a causal-masked transformer.

\subsubsection{Baseline Models.}
% We compare our method against SOTA LLM-enhanced baselines from two categories. (i) Generative Recommendation Methods: These models perform autoregressive generation for recommendation. We include \textbf{TIGER} \cite{tiger}, which uses an RQ-VAE and an encoder-decoder architecture, and \textbf{COBRA} \cite{cobra}, which employs a decoder-only architecture to fuse semantic IDs (from RQ-VAE) with dense behavioral vectors. (ii) Knowledge-Injected LLM Recommendation Methods: These models enhance LLMs with external or domain-specific knowledge. We select \textbf{UniSRec} \cite{unisrec}, which leverages contrastive pretraining for cross-scenario transfer; \textbf{TALLRec} \cite{tallrec}, which applies LoRA for few-shot recommendation; and \textbf{RecFormer} \cite{recformer}, which uses a two-stage, cross-domain pretraining and fine-tuning framework. Additionally, we also include \textbf{KAR} \cite{KAR}, which elicits world knowledge via user‑preference and item‑factual prompts, and \textbf{$\textbf{R}^4$ec} \cite{R4}, which performs iterative knowledge reflection by evaluating and updating retrieved knowledge before integration.
We first compare our method with SOTA prompt-based knowledge fusion methods: \textbf{KAR} \cite{KAR}, which elicits world knowledge via user‑preference and item‑factual prompts, and \textbf{$\textbf{R}^4$ec} \cite{R4}, which performs iterative knowledge reflection by evaluating and updating retrieved knowledge before integration. Second, we compare our method against SOTA LLM-enhanced baselines from two categories. (i) Generative Recommendation Methods: These models perform autoregressive generation for recommendation. We include \textbf{TIGER} \cite{tiger}, which uses an RQ-VAE and an encoder-decoder architecture, and \textbf{COBRA} \cite{cobra}, which employs a decoder-only architecture to fuse semantic IDs (from RQ-VAE) with dense behavioral vectors. (ii) Knowledge-Injected LLM Recommendation Methods: These models enhance LLMs with external or domain-specific knowledge. We select \textbf{UniSRec} \cite{unisrec}, which leverages contrastive pretraining for cross-scenario transfer; \textbf{TALLRec} \cite{tallrec}, which applies LoRA for few-shot recommendation; and \textbf{RecFormer} \cite{recformer}, which uses a two-stage, cross-domain pretraining and fine-tuning framework.

\subsubsection{Implementation Details.}
Our experiments are conducted on a distributed TensorFlow \cite{tensorflow2016abadi} platform with 2 parameter servers and 10 workers, each equipped with a single GPU. To ensure fairness, we use Qwen3-4.0B \cite{yang2025qwen3} as the base LLM to extract semantic knowledge for all methods, including our method and the baselines. For our CoCo framework, the embedding dimension is set to 128, and the hidden dimension of its transformer layers is 640. The soft prompt candidate pool in the collaboration phase is configured with a VQ codebook of size $K=64$, 2 public prompts, and a private prompt filtering threshold of $\theta=0.45$. In the contradiction phase, we employ LoRA for parameter-efficient fine-tuning \cite{hu2022lora} with a rank $r=8$, a scaling factor $\alpha_{lora}=16$, and a dropout rate of 0.05. The LoRA updates are applied to the query and value projection matrices within the attention layers. The loss weights are set as $\alpha=0.2$, $\beta=0.15$, and $\gamma=0.25$, respectively.
% We conduct our experiments on a distributed TensorFlow \cite{tensorflow2016abadi} platform consisting of two parameter servers and ten worker nodes, each equipped with a single GPU. To ensure a fair comparison, we maintain a consistent experimental setting. Specifically, all methods (our method and all baselines) use Qwen3-4.0B \cite{yang2025qwen3} to extract semantic knowledge. For the components of CoCo, the embedding dimension is 128 and the hidden dimension for its transformer layers is 640. The soft prompt candidate pool is configured with a VQ-VAE codebook size of $K=128$, 2 public prompts, and a private prompt filtering threshold of $\theta=0.45$. For the LLM fine-tuning stage, we employ LoRA with a rank $r=8$, a scaling factor $\alpha=16$, and a dropout rate of 0.05. These LoRA updates are applied to the query and value projection matrices within each attention layer for parameter-efficient adaptation \cite{hu2022lora}.

\subsection{Performance Comparison}
To answer RQ1, RQ2, and RQ3, we conduct a comprehensive comparison of CoCo's performance across 3 experimental scenarios: (1) comparison with knowledge fusion baselines across different backbones, (2) comparative analysis with early-stage LRSs, and (3) performance variation across different LLM scales.

\subsubsection{Comparison with knowledge fusion baselines across different backbones (RQ1).}
We first conduct comparative experiments between CoCo and state-of-the-art knowledge fusion-based LRSs on the Amazon dataset and a real-world e-commerce dataset using different behavioral backbones. The results are presented in Table \ref{table:backbone_experiment_results}. According to Table \ref{table:backbone_experiment_results}, we observe that:
\begin{itemize}[noitemsep, topsep=0pt, leftmargin=*]
    \item Our method achieves the best performance improvements across all backbones, with a maximum gain of 8.58\%. This result demonstrates that semantic knowledge extracted from LLMs can effectively complement the modeling of user intent in RSs, particularly in scenarios involving complex user behavior sequences.
    \item Compared to existing baselines, our method consistently achieves superior performance. This is due to two innovations: (1) the collaboration enhancement mechanism enables LLMs to dynamically generate semantic knowledge that is highly aligned with user behavior patterns; and (2) the introduced contradiction elimination mechanism preserves world knowledge while gradually aligning the LLMs' semantic space with the RSs' latent space.
    % \item Compared to existing baselines, our method consistently achieves superior performance. This is due to two innovations: (1) the collaboration enhancement mechanism enables LLMs to dynamically generate semantic knowledge aligned with user behavior patterns; and (2) the contradiction elimination mechanism preserves world knowledge while aligning the LLMs' semantic space with the recommendation system's latent space.
    \item Our framework demonstrates strong adaptability to any recommendation system architecture. This is because the semantic knowledge computation and alignment process does not change the model structure of RSs, and the LoRA fine-tuning introduces minimal parameter updates. Additionally, the contradiction elimination mechanism reduces computational overhead through conditional gradient freezing, enabling rapid integration into industrial-scale RSs.
\end{itemize}

\subsubsection{Comparative analysis with early-stage LRSs (RQ2).}
In this scenario, we compare the CoCo based on PinnerFormer with generative recommendation and early stage LRSs. The results are presented in Table \ref{table:baseline_experiment_results}. As shown in Table \ref{table:baseline_experiment_results}, CoCo significantly outperforms all baselines, achieving a maximum improvement of 5.03\% over the second-best model. Generative models (\textit{e.g.}, Tiger) rely on autoregressive token generation but lack access to world knowledge acquired during LLM pre-training, leading to generated outputs that often deviate from users' actual needs. Early LRSs inject world knowledge by using LLMs as encoders for multimodal features, resulting in unidirectional semantic knowledge flow that underutilizes LLMs' predictive reasoning capabilities. In contrast, our approach enables bidirectional optimization between LLMs and RSs through an end-to-end framework, preserving the integrity of world knowledge while fully leveraging LLMs' reasoning potential.

\subsubsection{Performance variation across different LLM scales (RQ3).}
To investigate the scalability of the CoCo framework, we replace its LLM components with different size models (Qwen3-0.8B/1.7B/4B/8B \cite{yang2025qwen3}), and the results are presented in Appendix \ref{appd::RQ3}. 

\subsection{Study on Hyperparameter Impact (RQ4)}
\subsubsection{Impact of $K$.}
\begin{figure}[htbp]
    \captionsetup{skip=2pt} 
  \includegraphics[width=0.48\textwidth]{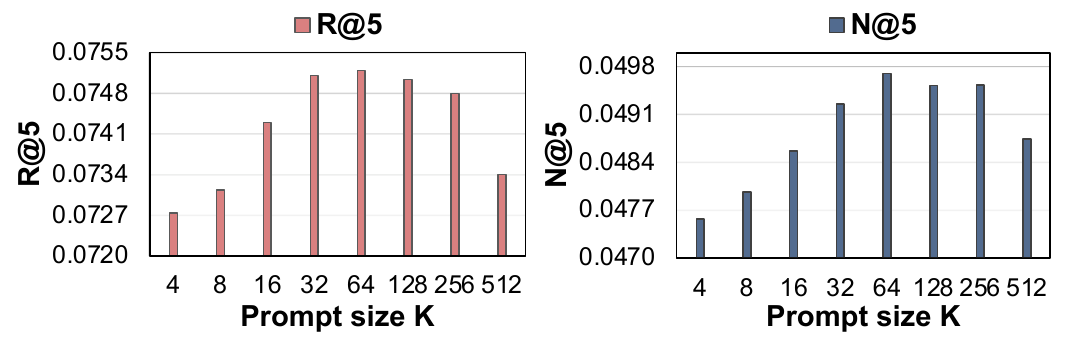}
  \caption{Sensitivity Analysis of the prompt candidate size $K$.}
  \label{fig:sensitive_k}
    \vspace{-10pt}
\end{figure}

We evaluate the impact of the prompt candidate size $K$ on real industrial datasets, as shown in Figure \ref{fig:sensitive_k}. The performance exhibits an initial significant improvement followed by a decline as $K$ increases from 4 to 512. When $K=4$, the marginal performance gain is limited, indicating that an extremely small candidate library fails to capture diverse user semantic needs. The performance reaches its peak at $K=64$, demonstrating that a moderately sized prompt library better models user intent through richer semantic representation. Beyond $K=128$, performance gradually declines due to increased semantic redundancy leading to information homogenization, as well as computational noise and convergence challenges from lower-quality prompts. In conclusion, we recommend setting $K=64$ to achieve an optimal balance between performance and computational efficiency. 
Furthermore, we visualize the distribution of $m$ when $K=64$ in the Appendix \ref{appd:m}.

\subsubsection{Impact of $\theta$.}
\begin{figure}[htbp]
  \includegraphics[width=0.48\textwidth]{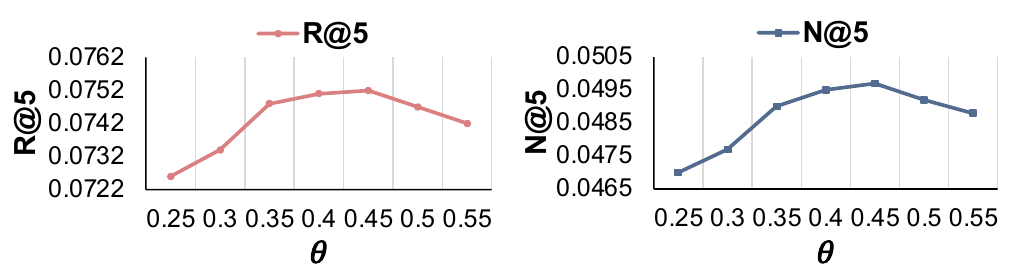}
 \vspace{-15pt}
    \captionsetup{skip=2pt} 
  \caption{Sensitivity Analysis of the threshold $\theta$}
  \label{fig:sensitive_theta}
    \vspace{-12pt}
\end{figure}

We investigate the impact of the prompt determination threshold $\theta$ on real industrial datasets, where $\theta$ governs the selection of user-specific private prompts. The results are presented in Figure \ref{fig:sensitive_theta}. When $\theta$ is set to a low value, the similarity requirement between user profiles and prompt semantics becomes lenient, resulting in excessively redundant or noisy prompts for each user. These low-quality prompts introduce noise that disrupts the recommendation system's decision-making process. As $\theta$ increases, the semantic consistency of private prompts improves significantly, leading to enhanced recommendation performance. However, overly stringent $\theta$ may cause performance degradation for some users due to insufficient available prompts that match their profiles.

\subsection{Analysis Experiments}
\subsubsection{Impact of Knowledge Fusion (RQ5).}
We evaluate the impact of different knowledge fusion methods in the alignment module on CoCo's performance and the results are presented in \ref{appd::RQ5}. 

\subsubsection{Visualization (RQ6).}
We visualize the behavioral and semantic spaces before and after applying CoCo in Appendix \ref{appd::RQ6}, demonstrating its effectiveness in integrating the two spaces.

\begin{table}[htbp]
\centering
\captionsetup{aboveskip=5pt,belowskip=5pt}
 \vspace{-5pt}
\caption{Ablation Study of CoCo}
\resizebox{0.8\linewidth}{!}{
    \begin{tabular}{c|cc|cc}
    \toprule
    \multirow{2}{*}{\textbf{Method}} & \multicolumn{2}{c|}{\textbf{Beauty}} &  \multicolumn{2}{c}{\textbf{Industrial Dataset}}\\
     & $R@5$ & $N@5$ &  $R@5$ & $N@5$  \\ 
    \midrule
    \small{CoCo} & \textbf{0.0549} & \textbf{0.0405} & \textbf{0.0752} & \textbf{0.0497} \\
    \small{$CoCo_{\text{Soft}}$} & 0.0531 & 0.0390 & 0.0737 & 0.0487 \\
    \small{$CoCo_{\text{Dec}}$} & 0.0539 & 0.0397 & 0.0738 & 0.0491 \\
    \small{$CoCo_{\text{Con}}$} & 0.0543 & 0.0400 & 0.0742 & 0.0493 \\
    \bottomrule
    \end{tabular}
}
\label{table:ablation_study_results}
\vspace{-12pt}
\end{table}

\subsection{Ablation Study (RQ7)}
In this section, we investigate the respective roles of the dynamic soft prompt generation module ($CoCo_{\text{Soft}}$), semantic decoupling module ($CoCo_{\text{Dec}}$), and contradiction elimination module ($CoCo_{\text{Con}}$).
\begin{itemize}[noitemsep, topsep=0pt, leftmargin=*]
    \item $CoCo_{\text{Soft}}$ denotes the use of fixed structured prompt inputs.
    \item $CoCo_{\text{Dec}}$ corresponds to removing the decoupling module and directly aligning LLM outputs with RSs outputs.
    \item $CoCo_{\text{Con}}$ corresponds to removing the contradiction elimination strategy and proceeding without fine-tuning the LLMs.
\end{itemize}
The ablation study results are presented in Table \ref{table:ablation_study_results}. Specifically, when using fixed semantic structured prompts, the N@5 metric on the Beauty dataset drops by 3.7\%, indicating that static prompts fail to adapt to users' personalized needs. Removing the semantic decoupling module leads to a 1.98\% decline in N@5, which can be attributed to increased semantic redundancy. 
Finally, if the LLMs' parameters are not updated, N@5 decreases by 1.23\%, demonstrating that static LLMs outputs fail to align with the recommendation system's latent space. The framework's performance gains arise from synergistic component interactions: the collaboration module generates dynamic prompts for personalized adaptation, the decoupling module eliminates redundancy via orthogonal constraints, and the contradiction module refines LLM outputs to align with the recommendation space.

\subsection{Online Experiments (RQ8)}
To further validate our approach, we conduct an online A/B test on an advertising recommendation platform of a leading e-commerce company in Southeast Asia from Sep 24 to 30, 2025. The control group is based on PinnerFormer, whereas the experimental group employs our proposed CoCo. Both groups consist of 10\% randomly selected users. Specifically, we observed a \textbf{1.91\%} increase in the \textbf{advertising revenue}, \textbf{0.64\%} increase in the \textbf{gross merchandise volume (GMV)} and \textbf{0.53\%} increase in the \textbf{click-through rate (CTR)}. These online results further validate the effectiveness of our proposed CoCo.
% The results of the online experiment once again confirm the performance of our proposed CoCo.

\begin{table}[htbp]
    \centering
    \caption{Results of the online A/B experiment, with all performance gains being statistically significant at $p < 0.05$.}
    \label{dataset2}
    \vspace{-10pt}  % 减少标题和表格之间的间距
    \resizebox{190pt}{!}{  % 缩放表格以适应双栏中的一栏宽度
        \begin{tabular}{c|ccc}
            \toprule
             method & Advertising Revenue & GMV  & CTR \\
            \midrule
            CoCo & +1.91\% & +0.64\% & +0.53\% \\
            \bottomrule
        \end{tabular}
    }
    \vspace{-15pt}
\end{table}
\section{Conclusion}
In this work, we propose CoCo, an end-to-end collaboration and contradiction fusion framework between large language models (LLMs) and recommender systems (RSs). We first elucidate the benefits of LLM-generated semantic knowledge for enhancing recommendations, while also addressing the inherent challenges in aligning disparate semantic and behavioral spaces. 
To tackle this, CoCo operates in two distinct phases: the collaboration enhancement phase dynamically generates highly personalized semantic knowledge tailored to individual user behavioral patterns.
Following this, the contradiction elimination phase adaptively adjusts the LLM's parameter distribution, enabling a deeper fusion of semantic and behavioral spaces.
Our framework can be integrated with any existing RS backbone without modifying the architecture or training procedures, offering a highly adaptable and practical solution for real-world recommendation.
% offering broad potential for practical application. 
In summary, CoCo presents a novel solution for leveraging personalized world knowledge from LLMs to enhance RSs, offering valuable insights for future research in this field.
Our research opens two crucial directions for future work: (1) Exploring larger-scale LLM deployment and novel scaling laws that can guide the fusion of LLMs and RSs. (2) Innovating fusion paradigms for the fundamental integration of LLMs and RSs.
\bibliographystyle{ACM-Reference-Format}
\bibliography{reference}

\clearpage
\newpage

\appendix

\section{Algorithm}
\label{appd:algorithm}
We present the algorithm of CoCo in Algorithm \ref{algo:1}.

\begin{algorithm}[H]

\caption{The Algorithm of CoCo Framework.}
\label{algo:1}
\begin{algorithmic}
\STATE \textbf{Input:} The recommendation model $f$, the unified feature vector $\mathbf{e}_{u}$ and LLM $f_L$.

\STATE \textbf{Parameters:} Codebook size $K$, semantic knowledge determination threshold $\theta$. 

\STATE \textbf{Output:} Optimized recommendation scores $ \hat{y} $.
\end{algorithmic}
\begin{algorithmic}[1]
   % \STATE Initialize Codebook $ C \leftarrow \emptyset $; \COMMENT{Universal Semantic Prompt}
   \STATE Initialize Codebook $ Z \leftarrow $ \COMMENT{Universal Semantic Prompts};
   
   \WHILE{not converged}
       \STATE Compute the similarity $s_i$ between $\mathbf{e}_{u}$ and all soft prompt vectors in the codebook $Z$;
       \STATE select the $m$ most relevant soft prompts with similarity scores exceeding $\theta$ for each user;
       \STATE Concatenate the prompt with the text-based behavior sequence embedding $[\mathbf{h}_{prompt}, \mathbf{h}_{text}]$ and input $f_L$;
       \STATE Compute quantization Loss $\mathcal{L}_Q$;
       % and commitment loss $\mathcal{L}_C$;
       \STATE Decouple the mixed semantic outputs of multiple soft prompts and LLMs and compute the orthogonal constraint loss $\mathcal{L}_{ortho}$;
       \STATE Compute the matrix $M$ and reconstruct the $f_L$ output;
       \STATE Align personalized knowledge and behavior vector $\operatorname{CrossAttention}\left(\mathbf{Q}=h_{RS}, \mathbf{K}=h_{concat}, \mathbf{V}=h_{concat}\right)$;
       \STATE Compute the final loss $\mathcal{L} = \mathcal{L}_r + \alpha \mathcal{L}_{aux}  + \beta  \mathcal{L}_{ortho} + \gamma \mathcal{L}_{Q}$, update the RSs, and fine-tune the LLM using LoRA.
   \ENDWHILE

\end{algorithmic}
\end{algorithm}

\section{Dataset Statistics}
\label{appd:dataset}
\begin{table}[htbp]
    \centering
    \caption{Statistics of Public and Industrial Datasets. \#Users and \#Items denote the number of users and items, respectively; \#Interactions denotes the total number of user–item interactions in each dataset.}
    \renewcommand{\arraystretch}{1.1}
    \label{fig:data_scale_comparison}
    \resizebox{230pt}{!}{  % 缩放表格以适应双栏中的一栏宽度
        \begin{tabular}{c|c|c|c}
            \toprule
            Dataset & Beauty & Toys & Industrial Dataset \\
            \midrule
            \#User & 22,363 & 35,598 & 18,631,691 \\
            \#Item & 12,101 & 18,357 & 23,572,371 \\
            \#Interaction & 296,175 & 167,526 & 3,438,193,075 \\
            \bottomrule
        \end{tabular}
    }
\end{table}

\begin{figure*}[ht]
  \centering
  \includegraphics[width=1.0\textwidth]{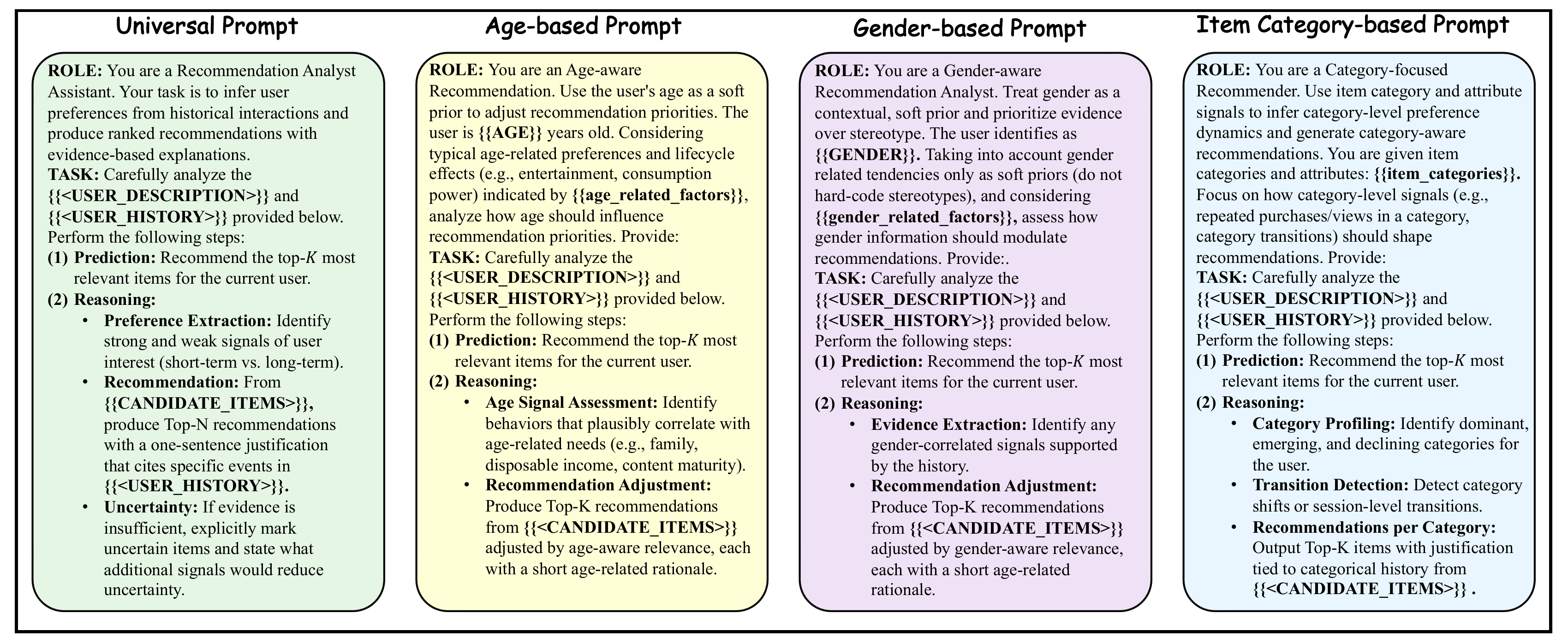 } % Reduce the figure size so that it is slightly narrower than the column.
  \caption{Four prompt templates used in the preliminary experiments. From left to right: general prompt, age-guided prompt, gender-guided prompt, and item category-guided prompt.}
  \label{fig:6}
\end{figure*}

Our evaluation is conducted on 3 distinct datasets: 2 derived from a public dataset and 1 large-scale industrial dataset. The statistics for each are summarized in Table~\ref{fig:data_scale_comparison}.
We construct two public datasets from the widely-used Amazon Product Reviews collection \cite{AmazonDataset}. Specifically, we select two categories with rich user interactions: "Beauty" and "Toys \& Games". For each user, we form an interaction sequence by chronologically ordering their reviews. Following common practice \cite{tiger}, we filter out users with fewer than five interactions to ensure data quality and meaningful sequence modeling. For evaluation, we employ a standard leave-one-out strategy \cite{sasrec}: for each user sequence, the most recent interaction serves as the test set, the second-to-last as the validation set, and all preceding interactions constitute the training set. 

The industrial dataset consists of recommendation logs from a Southeast Asian e‑commerce platform, covering 18 million users and 23 million ad records from January to July 2025, reflecting real user behavior and ad content. Its scale and complexity far exceed those of the public datasets, enabling a more realistic evaluation of system performance. In the industrial dataset, we extract a rich set of features. On the user side, these include (i) profile features (\textit{e.g.}, gender and location), (ii) behavioral features (\textit{e.g.}, click/conversion sequences from the past 180 days), and (iii) statistical features crucial for modeling interests \cite{fan2022modeling}. On the item side, we utilize (i) ID-based features (\textit{e.g.}, item ID, category ID, and seller ID) and (ii) historical statistical features.

\begin{table}[htbp]
% \centering
\captionsetup{aboveskip=5pt,belowskip=5pt}
\caption{Performance scaling of the CoCo method with larger LLMs. The "$Impr.$" column quantifies the relative performance improvement (\%) as model size increases, benchmarked against the smallest model (\textit{i.e.}, Qwen3-0.8B).}
\resizebox{0.8\linewidth}{!}{
    \begin{tabular}{c|cc|cc}
    \toprule
    \multirow{1}{*}{\centering LLM Version} 
    & $R@5$ & $Impr.$ &  $N@5$ & $Impr.$ \\
    \midrule
    Qwen3-0.8B & 0.0706 & — & 0.0478 & — \\
    Qwen3-1.7B & 0.0725 & $+2.69\%$ & 0.0486 & $+1.67\%$ \\
    Qwen3-4.0B & 0.0752 & $+6.52\%$ & 0.0497 & $+3.97\%$ \\
    Qwen3-8.0B & 0.0773 & $+9.49\%$ & 0.0503 & $+5.23\%$ \\
    \bottomrule
    \end{tabular}
}
% \vspace{-0.5cm}
\label{table:llm_version}
\end{table}

\section{Prompt Template}
We present the 4 prompt templates we used in the pilot experiments in Figure \ref{fig:6}.
\label{appd:pilot}

\section{Performance variation across different LLM scales (RQ3).}
\label{appd::RQ3}
To investigate the scalability of the CoCo framework, we replace its LLM components with models of different sizes (Qwen3-0.8B/1.7B/ 4B/8B) \cite{yang2025qwen3}, and the results are presented in Table \ref{table:llm_version}. As shown in Table \ref{table:llm_version}, the recommendation performance improves significantly as the parameter increases from 0.8B to 8B. Larger LLMs, through exposure to broader data during pre-training (\textit{e.g.}, cross-domain texts and user reviews), can generate semantic features that are more aligned with user intent. Additionally, large models demonstrate enhanced capabilities in understanding complex user behavior sequences, thereby improving knowledge relevance. Furthermore, the CoCo framework offers a novel perspective on the scaling laws of LRSs, demonstrating that scaling semantic knowledge can significantly enhance performance without requiring additional parameter counts or data scales in recommendation systems. However, we observe that in real-world massive data scenarios with latency and memory constraints, deploying extremely large LLMs becomes infeasible. This limitation will guide our future hardware optimization efforts.

\section{Impact of Knowledge Fusion (RQ5)}
\label{appd::RQ5}
We compare the base model, concatenation-based input, MLP-based fusion, and the CoCo framework on industrial datasets, as shown in Table \ref{table:fusion_comparison}. According to Table \ref{table:fusion_comparison}, two approaches that directly utilize LLM-derived semantic knowledge consistently result in performance degradation. This is primarily due to two factors: (1) direct concatenation of high-dimensional LLM embeddings (\textit{e.g.}, 896/4096 dimensions) with behavioral features leads to dimensionality explosion and (2) the noisy information in LLM outputs introduces non-convergence issues in recommendation models. In contrast, CoCo's cross attention based fusion mechanism enables adaptive selection of the most relevant semantic segments for each user, offering a semantic enhancement approach that avoids dimensionality explosion while preserving the core value of semantic knowledge.

\begin{table}[htbp]
\centering
\renewcommand{\arraystretch}{1.1} 
\captionsetup{aboveskip=5pt,belowskip=5pt}
\caption{Performance comparison of different knowledge fusion methods. The performance differences relative to the baseline (CoCo) are highlighted in bold.}
\resizebox{\linewidth}{!}{
    \begin{tabular}{c|cc|cc}
    \toprule
    \multirow{2}{*}{\textbf{Method}} 
    & \multicolumn{2}{c|}{\textbf{PinnerFormer}} 
    & \multicolumn{2}{c}{\textbf{SASRec}} \\
     & $R@5$  & $N@5$ & $R@5$  & $N@5$  \\
    \midrule
    CoCo
    & 0.0752 & 0.0497  & 0.0521  & 0.0355  \\
    $CoCo_{\text{mlp}}$ 
    & 0.0745 (\textbf{-0.93\%}) & 0.0494 (\textbf{-0.60\%}) & 0.0515 (\textbf{-1.15\%}) & 0.0352 (\textbf{-0.85\%}) \\
    $CoCo_{\text{fea}}$ 
    & 0.0742 (\textbf{-1.33\%}) & 0.0492 (\textbf{-1.01\%}) & 0.0513 (\textbf{-1.54\%}) & 0.0351 (\textbf{-1.13\%}) \\
    
    \bottomrule
    \end{tabular}
}
\label{table:fusion_comparison}
\end{table}

\section{Visualization (RQ6)}
\label{appd::RQ6}
\begin{figure}[hb]
\centering
\begin{minipage}[b]{0.48\columnwidth}
\centering
\includegraphics[width=\linewidth]{Figure4.pdf}
\end{minipage}\hfill
\begin{minipage}[b]{0.48\columnwidth}
\centering
\includegraphics[width=\linewidth]{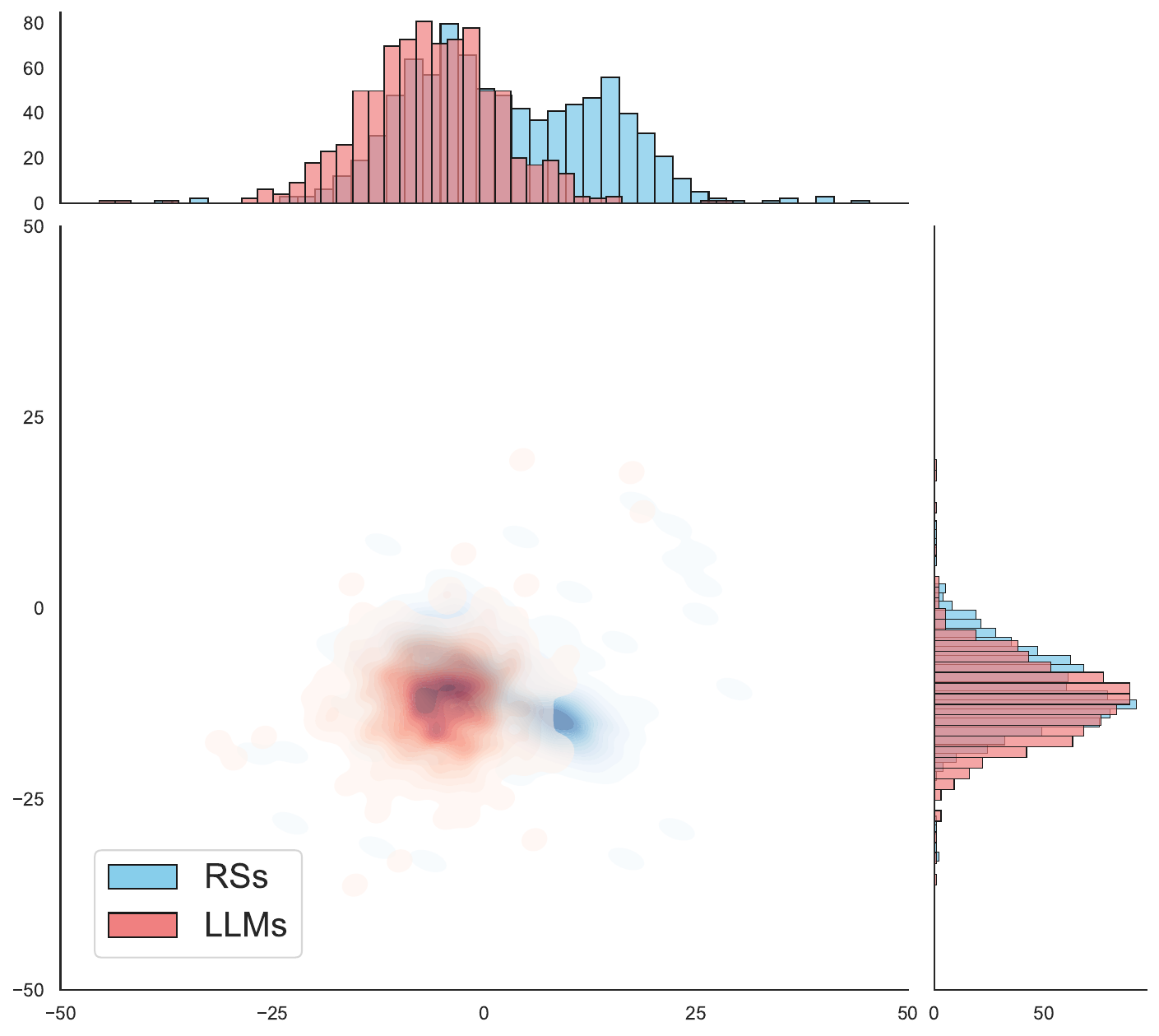}
\end{minipage}
\caption{Visualization of Behavioral and Semantic Spaces Before and After Applying CoCo (Left: Before; Right: After).}
\label{fig:visual}
\end{figure}

We randomly select 10000 LLMs output samples and RSs output samples, and visualize the distribution differences between behavioral space and semantic space, with and without the CoCo framework using t-SNE, as shown in \ref{fig:visual}. When the CoCo framework is not applied, the behavioral space and semantic space exhibit distinct separation: behavioral features form tight, concentrated clusters while semantic knowledge shows discrete, scattered distributions. In contrast, after introducing the CoCo framework, the two spaces gradually converge and cluster boundaries between behavioral features and semantic knowledge become blurred, with significantly increased overlapping regions. This phenomenon demonstrates that the CoCo framework effectively integrates the behavioral and semantic spaces through dynamic matching of semantic-behavioral features via cross attention mechanisms and continuous optimization of LLMs outputs through LoRA fine-tuning strategies.

\section{Visualization of $m$}
\label{appd:m}
\begin{figure}[htbp]
  \includegraphics[width=0.48\textwidth]{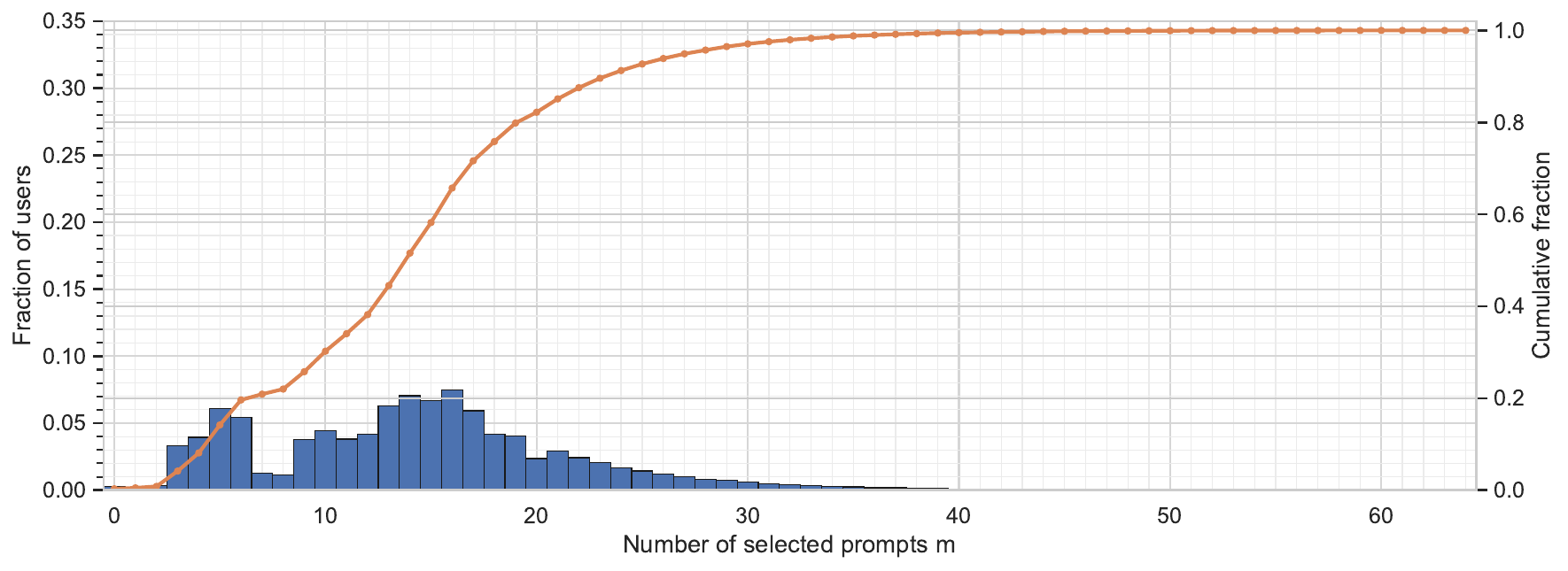}
  \caption{Visualization of the distribution of $m$.}
  \label{fig:m}
    % \vspace{-12pt}
\end{figure}

In Figure \ref{fig:m}, we visualize the distribution of $m$ when $K=64$. We find that most users' range are from 0 to 30.

\end{document}